\documentclass[epj]{svjour}
\pdfoutput=1
\usepackage{amsmath}

\newcommand{\beq}{\begin{equation}}
\newcommand{\eeq}{\end{equation}}

\newcommand{\appropto}{\mathrel{\vcenter{
  \offinterlineskip\halign{\hfil$##$\cr
    \propto\cr\noalign{\kern2pt}\sim\cr\noalign{\kern-2pt}}}}}

\usepackage{array, graphicx, subfigure}
\usepackage{graphics}

\begin{document}
\title{  Using Drell-Yan forward-backward asymmetry to reduce PDF uncertainties in the measurement of electroweak parameters}
\subtitle{  }
\author{{\bf A.  Bodek,  J. Han, A. Khukhunaishvili, W. Sakumoto}}
\institute{Department of Physics and Astronomy, University of
Rochester, Rochester, NY  14627-0171
}
\date{Received: date / Revised version: Feb 18, 2016,~~~~~~~~~ {\bf  ArXiv:1507.02470v3 [hep-ex] }}
%
\abstract{
The uncertainties in Parton Distribution Functions (PDFs)  are the dominant source of the  systematic uncertainty in precision measurements of electroweak parameters at hadron colliders (e.g.  $\sin^2\theta_{eff}(M_Z)$, $\sin^2\theta_{W}=1-M_W^2/M_Z^2$ and the mass of the W boson).  We show that measurements of the  forward-backward charge asymmetry ($A_{FB}(M,y)$)  of Drell-Yan  dilepton events produced at hadron colliders   provide a new powerful tool to  reduce the PDF uncertainties in these measurements.
} 
\titlerunning{ Using Drell-Yan $A_{FB}$ to reduce PDF uncertainties}
\maketitle

\section {Introduction}
Precision measurements in hadron colliders are  limited by our knowledge of Parton Distribution Functions (PDFs).
In general, PDF fits by various groups including  \textsc{cteq} \cite{CTEQ}, \textsc{mmht}\cite{MMHT}, \textsc{nnpdf} \cite{NNPDF,NNPDF2}, \textsc{hera}\cite{other}, and  \textsc{abm}\cite{ABM} are extracted from fixed target experiments and various cross sections
measurements at colliders.  The fixed target  experiments include  electron, muon, neutrino, and  Drell-Yan experiments. The collider experiments include $ep$(HERA),  ${\bar p}p$ (Tevatron)  and $pp$(LHC). 

Some of the fixed target measurements are on nuclear targets resulting in additional uncertainties from modeling 
of nuclear effects. Some  of the fixed target measurements are also at   low momentum transfers where the  contributions
of  non-perturbative and higher twist effects may be significant. 
These issues are absent in collider cross section data. Therefore,  recent PDF fits
have placed a greater emphasis on  collider cross section data. 

 

    \subsection {Measurements of electroweak parameters at hadron colliders}
    
     Within the standard model, measurements of the mass of the $Z$ boson and top quark, in combination with the
    mass of the Higgs boson,  can be used to predict  the mass of the W boson ($M_W$).
   At present,  the average of the all direct measurements of $M_W$
    (80385$\pm$15 MeV) is about 1.5 standard deviation higher\cite{pdgSM} than the prediction of the standard
    model.  Predictions of  supersymmetric models
     for $M_W$ are also higher than the predictions of the standard model\cite{Wmass}. 
     Therefore, more precise measurements of the mass of $M_W$
    are of great  interest.
     
     Alternatively, $M_W$ can also be extracted indirectly from
     measurements of the on-shell electroweak mixing angle $\sin^2\theta_{W}$ 
    by  the  relation $\sin^2\theta_{W}=1-M_W^2/M_Z^2$.  
      
      Measurements of the  forward-backward charge asymmetry in Drell-Yan
         dilepton events produced at hadron colliders 
(in the region of the $Z$ pole)   have been used to measure
 the value of the  {\it effective} electroweak (EW)  mixing
  angle $\sin^2\theta_{eff}^{lept} (M_Z)$\cite {cdf-ee,cdf-mumu,Dzero,ATLAS}.
   In addition, by incorporating electroweak radiative corrections in the analysis
   the CDF collaboration has also measured the  {\it on-shell} ~EW mixing angle  $\sin^2\theta_W$\cite{cdf-ee,cdf-mumu}.
 
        An uncertainty of $\pm$0.00030 in the measurement of $\sin^2\theta_{W}$ 
      is equivalent to an indirect measurement  of $M_W$ to a precision of $\pm$15 MeV.  However,  the PDF
      uncertainty quoted in the most recent measurement of $\sin^2\theta_{eff}$ by the ATLAS collaboration\cite{ATLAS}
      at the LHC is $\pm$0.00090.  Therefore, a significant reduction in the PDF uncertainty is needed.
   In this communication, we show  how $A_{FB}$  data also provide a new powerful tool
  to reduce  PDF uncertainties in the measurements of electroweak
parameters in hadron colliders 

    The constraints provided by $A_{FB}$ measurements in combination with constraints from the W charge asymmetry ($A_W$) 
     can  be used to reduce  the  PDF uncertainty in the extracted value of $\sin^2\theta_W$ and $\sin^2\theta_{eff}^{lept} (M_Z)$
     from $A_{FB}$  data.
   The  $A_{FB}$ constraints on PDFs can also be used to reduce the PDF uncertainty in other precision measurements with $Z$ and $W$ bosons such as the measurement  of $W_W$. 
   
   Asymmetries such as $A_{FB}$ and $A_W$ are ideal in providing additional constraints because asymmetries 
   are less sensitive to the choice of  QCD scale and QCD higher order terms.  In addition, there are  new techniques that  can be used\cite{scale,weighting} to greatly reduce the experimental systematic uncertainty in asymmetry measurements.

   \section{ $q\bar{q}$ annihilations to dileptons}
   In leading order (LO) dileptons are primarily produced in  quark-antiquark annihilation.  Here, 
one parton (quark or antiquark) carries momentum $x_1$ and another parton carries momentum $x_2$.
The momentum fractions  $x_{1,2}$ carried by the partons
 are related to the mass (M) and rapidity (y) of the two leptons as follows:
\begin{eqnarray}
x_{1,2} &=& \frac{M}{\sqrt{s} }e^{\pm y}
\label{eq1}
\end{eqnarray}
The angular dependence of the differential cross section for
$q\bar{q}$ annihilation to a dilepton pair can be written as
\begin{eqnarray}
\frac{d\sigma (M)}{d( \cos\theta )} &\propto & (1 + \cos^{2} \theta) + A_4(M)
\cos\theta 
\label{eq2}
\end{eqnarray}
\noindent  where  $\theta$ is the emission angle of the negatively charged 
lepton relative to the quark momentum in the dilepton center of mass frame,
and $A_4 (M)$ is  parameter that depend on the weak isospin and
charge of the incoming quarks. 

The cross sections for forward ($\sigma_{F}$) and
backward ($\sigma_{B}$) events  are given by

\begin{eqnarray}
\label{eq3}
\sigma_{F}(M) &= &\int_{0}^{1} \frac{d\sigma}{d(\cos\theta)}
d(\cos\theta) \\\nonumber
& \propto & \left(1+\frac{1}{3}\right) +  A_4(M) \left(\frac{1}{2}\right) \\
\label{eq4}
\sigma_{B}(M) &=& \int_{-1}^{0} \frac{d\sigma}{d(\cos\theta)}
d(\cos\theta) \\\nonumber
& \propto &  \left(1+\frac{1}{3}\right) -  A_4(M)\left(\frac{1}{2}\right)
\nonumber
\end{eqnarray}

The electroweak interaction introduces an asymmetry (a linear
dependence on $\cos\theta$), which can be expressed as
\begin{eqnarray}
\label{eq5}
A_{FB} (M)& =&  \frac{\sigma_{F}-\sigma_{B}}
{\sigma_{F}+\sigma_{B}} = \frac{3}{8}A_4(M)
\end{eqnarray}
The dependence of $ A_{FB} (M,y)$ on  $\sin^2\theta_{eff}^{lept}$
has been used to measure $\sin^2\theta_{eff}^{lept}$
at the  Tevatron and LHC.

 The systematic uncertainties in the measurement of  $ A_{FB} (M,y)$ 
 can be greatly reduced if $ A_{FB} (M,y)$
is extracted from a measurement of   $A_4(M,y)$.
This can done by 
an event weighting technique\cite{weighting} for which
there is a  cancelation of 
systematic errors that originate from 
 uncertainties in  acceptance and efficiencies. With this technique, 
  no acceptance or efficiency corrections are needed.
The extracted values of $A_4(M,y)$ using the event weighting technique
are  equal to the Born level  $A_4(M,y)$. This technique has been using
in the most recent measurements at CDF\cite{cdf-mumu}. 
 
\begin{figure}[ht]
\includegraphics[width=7.5 cm, height=6.0 cm]{{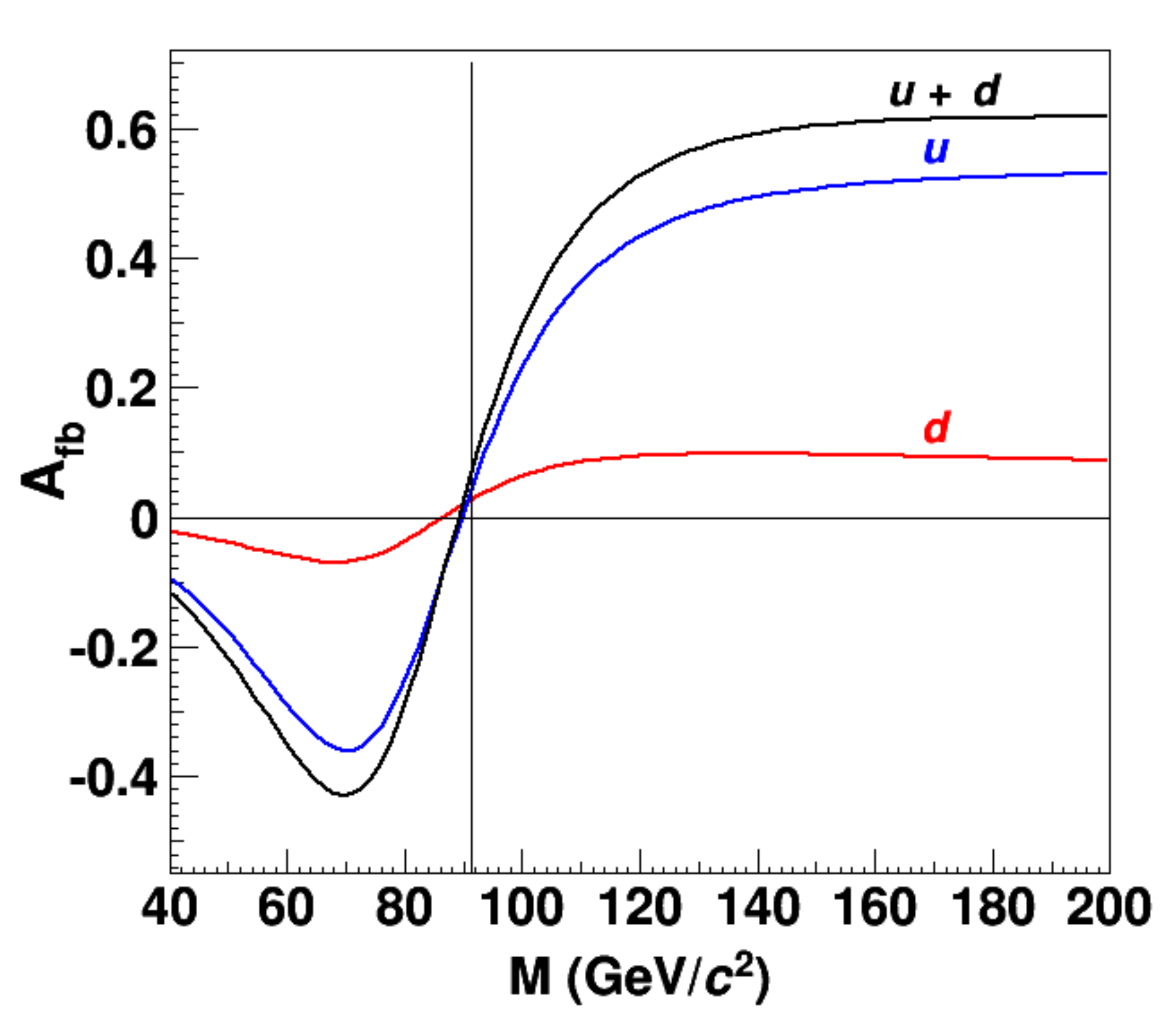}}
\caption{  The contributions of 
  u-type quarks   (blue) and  d-type quarks (red)
to $A_{FB}(M)$ at the Tevatron.}
\label{fig_1}
\end{figure}
\section { $A_{FB}$ at the Tevatron}
%
For $\bar{p}p$ collisions, the direction of
the quark is predominately in the proton direction, and  the
direction of the antiquark is predominately in the antiproton
direction. Here,   most of the cross section
originates from the annihilation of quarks in the proton
with antiquarks in the antiproton. Therefore, 
$A_{FB}$ is measured under the assumption that the quarks
originate form the proton, and the antiquarks originate from the 
antiproton (first term in eq.{\ref{eq6}). 

Since $q(x)$ in the proton is equal to ${\bar{q}}(x)$ in the antiproton,
the dilepton production cross section can be expressed as follows:
\begin{eqnarray}
\label{eq6}
\frac{d \sigma}{dM}(\bar{p}p) \propto  { \sum_{flavor}{ v_{i}
\{q_i(x_{1})\cdot{{q}_i(x_{2})}+{\bar q}_i(x_{1})\cdot{\bar{q}_i(x_{2})}} \} }
\end{eqnarray} 

Here $q_i(x)$ denote the quark distributions  ($u(x)$,
$d(x)$, $s(x)$ , $c(x)$,  $b(x)$)  and  $\bar{q}_i(x)$ denotes\ the
antiquark distributions ($\bar{u}(x)$,  $\bar{d}(x)$,
  $\bar{s}(x)$,   $\bar{c}(x)$,  $\bar{b}(x)$)  in the nucleon. 
   The parameters $v_{i}$ denote the
$Z/\gamma$  couplings for each flavor.  Here, 
$v_{i}$ are functions of both 
the dilepton mass and  $\sin^2\theta_{eff}^{lept}$.

The extraction of $\sin^2\theta_{eff}^{lept}$ from
$A_{FB}(M)$  (or $A_4(M)$)  is sensitive to PDFs
for two reasons.  First, $A_{FB}(M)$ for
charge 2/3 (u-type) quarks 
and   charge 1/3 (d-type)  quarks is different. Fig. \ref{fig_1} shows
 the contributions of   u-type quarks (blue),  d-type  quarks (red)
and the sum of the two contributions (black) to $A_{FB}(M)$ at the Tevatron
as given by
\begin{eqnarray}
\label{eqdu}
A_{FB}^{d-type}  &\approx&   \frac {(dd)_F-(dd)_B} {  (dd)_F+(dd)_B + (uu)_F+(uu)_B }\nonumber\\
A_{FB}^{u-type } &\approx&   \frac {(uu)_F-(uu)_B} {  (dd)_F+(dd)_B + (uu)_F+(uu)_B }\nonumber
\end{eqnarray} 
%
  The measured  asymmetry 
 is sensitive to the fraction of down quarks
in the proton because the  asymmetries
  for up and down quarks are different. The sensitivity is proportional to 
\begin{equation}
D^{Tev}_{AFB}(d)  \appropto \frac {d( x_1) ~ d(x_2) } {u (x_1) ~  u (x_2) } =[\frac{d}{u}(x_1)] [\frac{d}{u}(x_2)].
\label{eq7}
\end{equation}
In addition, there is a  small fraction of events for
which the annihilation is between sea antiquarks in the
proton with a sea quarks in the antiproton (second term in eq. \ref{eq6}). 
The forward-backward asymmetry $A_{FB}(M)$ of the second term
in equation \ref{eq6} is opposite to the $A_{FB}(M)$ of the larger first term. 
This also results in a  dilution ($D^{Tev}_{AFB}(\bar{q})$) of  the
measured asymmetry. 
\begin{eqnarray}
\label{eq8}
D^{Tev}_{AFB}(\bar{q})  \propto \frac { 
{  \sum_{flavor}{v_{q}{\bar q}(x_{1})\cdot{\bar{q}(x_{2})}}  }
}
{{u(x_1) {u}(x_2)}}
\end{eqnarray} 


%
%
The antiquark dilution is primarily from
 $u$ type antiquarks. 
For proton-antiproton collisions,  most of the cross
section is near $y$=0 ($x_1\approx x_2$).
Therefore, 
the PDF uncertainty in the extraction of $\sin^2\theta_{eff}^{lept}$ from
$A_{FB}(M)$  (or $A_4(M)$)  at the Tevatron depends primarily on
how well we can constrain the following contributions
to the dilution at  $x_{1} = M_z/\sqrt{s}$.

\begin{eqnarray}
\label{eq9}
D ^{Tev}_{AFB}(d)  &\appropto& [\frac{d}{u} (x_1)]^2\\
D ^{Tev}_{AFB}(\bar{q}) &\appropto& [\frac{\bar{u}}{u}(x_1)]^2
\end{eqnarray}

\subsection {$W$ charge asymmetry at the Tevatron}

The $W^-/W^+$ ratio  at the Tevatron can be written as 
\begin{equation}
(\frac{W^-}{W^+})^{Tev} \approx  \frac {d( x_1) ~ u(x_2) +s( x_1) ~c(x_2)  } {u (x_1) ~  d (x_2) +c( x_1) ~ s(x_2)  } \approx  \frac{d}{u}(x_1) /\frac{d}{u}(x_2)
\end{equation}

Precise measurements of
the $W$ asymmetry provide information on the $d/u$ ratio at the Tevatron.
  These measurements
are important to constrain the PDF uncertainties for the direct
measurement of the W mass.
 However, at the Tevatron
these measurements do not provide  information relevant to the  
measurement of  $\sin^2\theta_{eff}$ for two reasons.  
First, there is no information at   $y$=0 ($x_1\approx x_2$) since
here the $W$ charge asymmetry at the Tevatron is zero. Secondly, at the Tevatron, 
the $W$ charge asymmetry does not provide information
on the absolute level of $\frac{d}{u} (x)$. The $W$ charge asymmetry
at the Tevatron provides information  only on the slope of
$\frac{d}{u} (x)$ as a function of  $x$.

\subsection{PDF uncertainties: Hessian and Replica PDFs}
All  PDF groups provide a default (central) PDF set.
There are two methods that are used for the
determination of PDF uncertainties.  The first
method is to provide a set of eigenvector
error PDFs (Hessian  method). 
The PDF uncertainties in a measurement are
determined by repeating the analysis for
all of the error PDF sets, and adding in quadrature
the difference in the results obtained with  the  error
PDFs and the results obtained  with the  default
PDF.

The second method (which is referred to as replica PDFs) is to
provide a set  of  N (e.g. 100 or 1000) 
replica PDFs. Each of the PDF replicas
has equal probability of being correct.  The
central value of any observable 
is the average of the values  $s_i$= $(\sin^2\theta_W)_i$ 
extracted with each one of the N 
PDF replicas.  The PDF uncertainty (=$\sigma_{pdf}$) is the \textsc{rms}  of the
values extracted using all  N replicas. 
\begin{eqnarray}
\langle s \rangle &=&\frac {1}{N}  \sum_{i=1}^{N}{s_i}\\
\sigma_{pdf}&=&\sqrt { \frac {  \sum_{i=1}^{N}{(s_i-\langle s \rangle )^2}}{N-1}}
\end{eqnarray}
and the uncertainty in the estimate of the  PDF uncertainty is
$\Delta \sigma_{pdf}= \frac {\sigma_{pdf}}{\sqrt{2(N-1)}}$

The two methods provide equivalent information.
 For any given a set of Hessian
eigenvector PDFs there is a prescription to generate\cite{NNPDF2,weights-MST,weights-web}  an arbitrary
number of PDF replicas.
\subsection{Reducing PDF uncertainties with new data}
The advantage of the PDF replica method is that
constraints from new data can easily be incorporated
in any analysis by applying  different weights
for each replica.

Replicas for which the theory predictions 
are in agreement with the new data are given higher
weights, and replicas for which the predictions 
are in poor agreement are given lower weights.
The weights are derived from the $\chi^2$
values of the comparison between the new
data and theory prediction  
each of the PDF  replicas.

 The central value of any observable 
is  the {\it weighted}  average of the values
extracted using  each one of the N 
PDF replicas.  The PDF uncertainty is the  {\it weighted}  \textsc{rms} 
(root mean square)  of the
values extracted  each of the N replicas. 

The procedure of including constraints from
new data was initially proposed by   Giele and Keller\cite{GK}.
They proposed that  each of the N PDF replicas be weighted as follows:
\begin{eqnarray}
w_i&=&\frac {~ e^ {-\frac{1}{2}\chi^2_i}}{
\sum_{i=1}^{N}{~ e^ {-\frac{1}{2}\chi^2_i}}}\\
\label {GKw}
\langle s \rangle &=& \sum_{i=1}^{N}{w_i s_i}\\
\sigma_{pdf}&=&\sqrt { \frac {  \sum_{i=1}^{N}{w_i(s_i-\langle s \rangle )^2}}{1-1/N_{eff}}}
\end{eqnarray}

%
The weights reduce the effective number of replicas from N to $N_{eff}$ where
\begin{equation}
N_{eff}= \frac{1}{\sum_{i=1}^{N}{w_i^2}}
\label {effective}
\end{equation}
and the   uncertainty in the  estimate of the PDF uncertainty is
$\Delta \sigma_{pdf}\approx \frac {\sigma_{pdf}}{\sqrt{2(N_{eff}-1)}}$.

More recent discussions of the method can be found in references
\cite{weights-MST,weights-web,GK1,GK2,Ball}.
In the sections that follow we show how the mass and rapidity dependence of
$A_{FB}$ can be  used to both provide additional constraints
and reduce the PDF uncertainty in measurements  of $ \sin^2 \theta_W $.

\subsection{Number of replicas needed} 
Typically between 100 and 1000 PDF replicas are used.  A large number of replicas is only
needed if the new data that is being incorporated is so precise that the number of
effective replicas drops below 10.  This only happens if the statistical errors of the
new data are much smaller than the PDF uncertainties. 

 For the electroweak measurements that are discussed in this paper 
 the statistical errors which are achievable in the next few years are typically
 within a factor of 2-3  of the PDF uncertainties. Therefore,   100 replicas are typically sufficient.

\subsection{Mass dependance of $A_{FB}(M)$ as a function of  $\sin^2\theta_W$ and PDFs at the Tevatron}

The sensitivity of  the mass dependence  of $A_{FB}(M)$ on $\sin^2\theta_W$
 and  PDFs is different.    In the region of the $Z$ pole,   $A_{FB}(M)$ 
 is sensitive to the vector couplings, which depend on   $\sin^2\theta_W$.
 At higher and lower mass  $A_{FB}(M)$ is sensitive to the
 axial coupling and therefore insensitive to value of  $\sin^2\theta_W$.
 
 In contrast,  the magnitude of the  dilution of  $A_{FB}(M)$ depends on  the PDFs.
  The sensitivity to PDFs is largest in regions where   $A_{FB}(M)$ is large
  (i.e. away from the $Z$ pole). 
  %
  \begin{figure}
\includegraphics[width=8. cm, height=5.0 cm]{{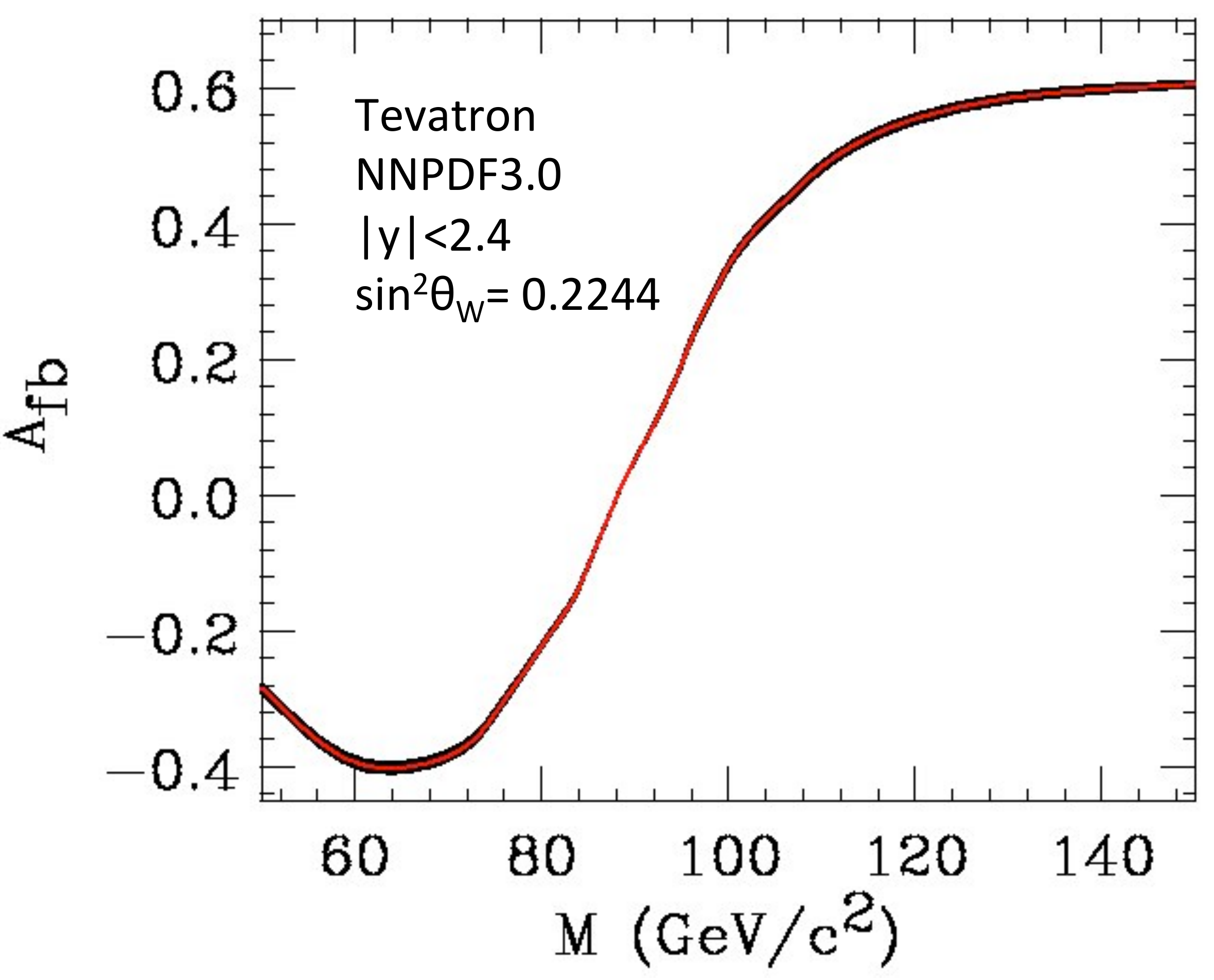}}
\caption{ $A_{FB}$ 
versus dilepton mass at the Tevatron  for  $ \sin^2 \theta_W $=0.2244 
and the default \textsc{nnpdf} 3.0 (\textsc{nnlo})  PDF (261000). The band corresponds to ten \textsc{nnpdf}  replicas.}
\label{fig_2}
\end{figure}
%

Fig.~\ref{fig_2} shows $A_{FB}(M)$ as a function 
dilepton mass at the Tevatron  for $\sin^2\theta_W$=0.2244.
 The band corresponds to the predicted
 values of $A_{FB}(M)$ for the default \textsc{nnpdf} 3.0 (\textsc{nnlo})  PDF (261000),
 and ten \textsc{nnpdf} 3.0 (\textsc{nnlo}) replicas.
$A_{FB}(M)$ is shown for
$\sqrt {s}$=1.96 TeV and
dilepton rapidity less 1.7, which corresponds to a typical acceptance for Tevatron experiments
(CDF or D0).

Fig.~\ref{fig_3}(a) shows the sensitivity of  $A_{FB}(M)$ 
at the Tevatron to PDFs.
The lines are the  difference between $A_{FB}(M)$ 
for 10 \textsc{nnpdf} 3.0 (\textsc{nnlo}) replicas and $A_{FB}(M)$ calculated
for  the central default \textsc{nnpdf} 3.0 (\textsc{nnlo}) (261000). 
Here $\sin^2\theta_W$ is fixed at a value of 0.2244. The difference
originates from the  differences in 
$\frac{d}{u} (x)$ and 
the antiquark fractions
 for the different PDF replicas.


Fig.~\ref{fig_3}(b) shows the sensitivity of  $A_{FB}(M)$ 
at the Tevatron to $\sin^2\theta_W$. 
The lines are the  difference between the calculated 
 $A_{FB}(M)$  for  $\sin^2\theta_W$  values ranging
from 0.2220 (show at the top in red)  to 0.2265 
(shown in the bottom in blue) and  $A_{FB}(M)$
for $\sin^2\theta_W$=0.2244.   Here  $A_{FB}(M)$ is calculated
with  the default \textsc{nnpdf} 3.0 (\textsc{nnlo}) (261000). 
%

\begin{figure}[ht]
\includegraphics[width=8.5 cm, height=9.0 cm]{{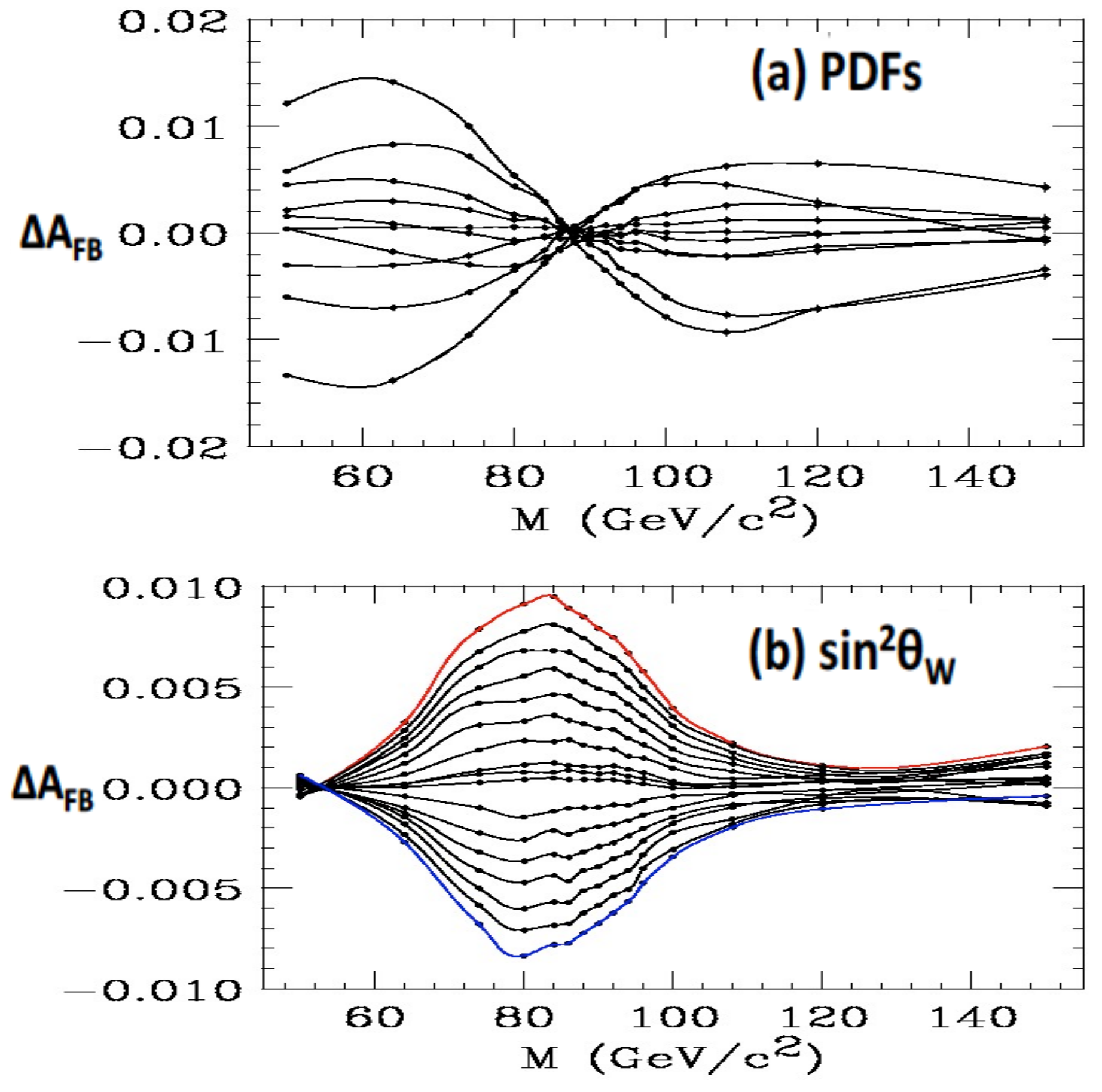}}
\caption{
Tevatron: (a) The  difference between $A_{FB}(M)$ 
for 10 \textsc{nnpdf} 3.0 (\textsc{nnlo}) replicas and $A_{FB}(M)$ calculated
for  the  default \textsc{nnpdf} 3.0 (\textsc{nnlo}) (261000). Much of the difference
originates form the  different dilution factors for each of the \textsc{nnpdf}   replicas.
Here $\sin^2\theta_W$ is fixed at a value of 0.2244. (b) 
The  difference between 
$A_{FB}(M)$  for different values of $\sin^2\theta_W$ ranging
from 0.2220 (shown at the top in red)  to 0.2265 (shown on the
bottom in blue), and 
$A_{FB}(M)$ for $\sin^2\theta_W$=0.2244.
 Here  $A_{FB}(M)$ is  calculated
with  the default \textsc{nnpdf} 3.0 (\textsc{nnlo}). }
\label{fig_3}
\end{figure}

As shown in Fig.~\ref{fig_3}(a)  there 
is a large difference in the  $A_{FB}(M)$
predictions for PDF sets with different 
$\frac{d}{u} (x)$ and antiquark
fractions $\frac{\bar{q}}{q}(x)$
in  regions where  $A_{FB}(M)$
is large and positive  (M$>$100 GeV).
 The changes in $A_{FB}(M)$ in  
regions where  $A_{FB}(M)$ is large and negative
(M$<$80 GeV) are in  the opposite direction.

In contrast, as shown in Fig.~\ref{fig_3}(b), different
values of  $\sin^2\theta_W$ change
$A_{FB}(M)$ primarily in the region  
near the $Z$ pole.  However, here the change
is in the same direction above and below the
$Z$ pole.  
Therefore, if we extract  $\sin^2\theta_W$ 
from $A_{FB}(M)$  data  with  different PDFs, 
PDFs with poor values of $\chi^2$ are less likely to be
correct.

\subsection{MC studies of dilepton production at  Tevatron }
The 10 fb$^{-1}$ Run II  $e^+e^-$ data sample at
CDF corresponds to about 500K events. A similar sample 
was collected by the D0 experiment\cite{Dzero}.  The acceptance of the Tevatron
experiments limits the sample to events with dilepton rapidity $|y|<$1.7.

We simulate $A_{FB}(M)$  measurements 
corresponding a  10 fb$^{-1}$ statistical sample
at the Tevatron  with three different input assumptions for $A_{FB}$.
In all cases we use   $\sin^2\theta_W$=0.2244 and  calculate
$A_{FB}$ in 15 bins for dilepton mass
spanning the range from M=50 GeV to M=150 GeV.
We generate pseudo data for three  input assumptions.
For each input assumption we generate 
a set of 1600 pseudo-experiments.

\begin{itemize}
\item The input  assumption for the first set of 1600 pseudo experiments
is that  $A_{FB}(M)$ is equal
 to the predictions of a  Tree-level calculation (including EBA EW radiative
corrections\cite{cdf-ee,cdf-mumu}) calculated with  the default  \textsc{nnpdf} 3.0 (\textsc{nnlo}) 
PDF set. 

\item The input  assumption for the second set of 1600
 pseudo experiments is that  $A_{FB}(M)$ is equal
 to the predictions of a  Tree-level calculation (including EBA EW radiative
corrections\cite{cdf-ee,cdf-mumu})   calculated with the  default  \textsc{nnpdf} 2.3 (\textsc{nnlo}) 
PDF set. 

\item The input  assumption for the third set of 1600
 pseudo experiments is that  
$A_{FB}(M)$ is equal  to the predictions of \textsc{resbos} \cite{ResBos}
(modified to include EBA EW radiative corrections\cite{cdf-ee,cdf-mumu}) 
calculated with the \textsc{cteq} 6.6 PDF set.
\end{itemize}

\begin{figure}[ht]
\includegraphics[width=8.5 cm, height=6.0 cm]{{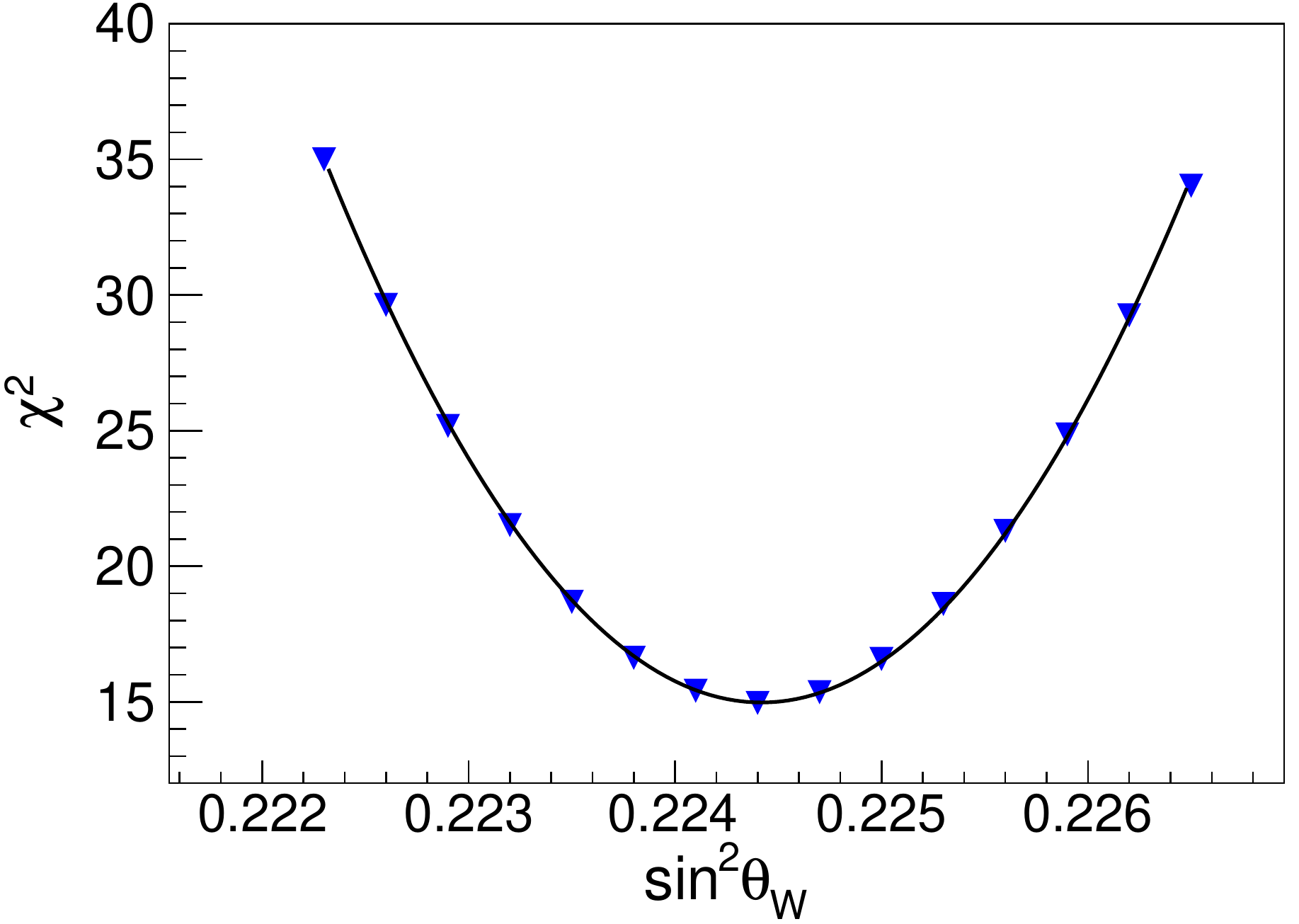}}
\caption{
 An example of the extraction of $\sin^2\theta_W$ from $A_{FB}(M)$  
data at the Tevatron.
 Here,  $\chi^{2}_{Afb}$
is plotted for different values of  $\sin^2\theta_W$ 
The extracted value of
$\sin^2\theta_W$ is the value with the minimum $\chi^{2}_{Afb}$
and the statistical error corresponds to a change of
$\chi^{2}_{Afb}$ by $\pm$1.  }
\label{fig_4}
\end{figure}
\begin{table}[htb]
\caption {Values of  $\sin^2\theta_W$ 
 with statistical errors and  PDF uncertainties  
 expected at the Tevatron for a
 10 fb$^{-1}$ sample for a CDF like detector.   The 
PDF uncertainty for a standard  analysis is compared to the PDF
uncertainty for an analysis with $\chi^{2}_{Afb}$ weighting.
The default  \textsc{nnpdf} 3.0 (\textsc{nnlo}) is used to generate the pseudo data
in the first column and 
 the  default \textsc{nnpdf} 2.3 (\textsc{nnlo}) is used
to generate the pseudo data in the second column.
All pseudo data are generated with $\sin^2\theta_W$=0.22420.
}
    \begin{center}
\begin{tabular}{|c||c|c|c||}
\hline
	&	Input	&  Input		\\
	CDF-like~detector	&	Tree-level	&Tree-level		\\
   Pseudo-Experiment&	Default	&	Default	\\
{{$Tevatron~10~fb^{-1}$}}   & {\textsc{nnpdf} 3.0 } &	{\textsc{nnpdf} 2.3 }   \\
$500K$ reconstructed &	(261000) & (261000)	\\
$e^+e^-$ events &(\textsc{nnlo})	 & (\textsc{nnlo}) 	\\
\hline
$\sin^2\theta_W$ input &	0.22420& 0.22420 	\\
\hline\hline
{statistical~error}	&	$\pm$0.00042	&	$\pm$0.00042	\\
$\Delta\sin^2\theta_W$		&		&		\\
CT10 PDF~error  	&		$\pm$0.00026	&	$\pm$0.00026		\\
\hline \hline
Number of analysis replicas	&100		&	100	\\ 
\textsc{nnpdf}  replica set 	&\textsc{nnpdf} 3.0 	&\textsc{nnpdf} 2.3 		\\ 
Templates		&	Tree-level 	&Tree-level		\\
		&		&		\\ 
Average  method &	$N_{eff}=100$			&$N_{eff}=100$		\\
extracted ~$\sin^2\theta_W$		&	0.22420	&0.22420		\\ 
PDF~error~\textsc{rms}   	&		$\pm$0.00027	&	$\pm$0.00028		\\
(uncertainty in PDF error)&$(~~0.00002)$		&$(~~0.00002)$		\\ 
&		&		\\ 
 $\chi^{2}_{Afb}$~weighting		&	$N_{eff}=88$			&$N_{eff}=85
 $		\\
extracted ~$\sin^2\theta_W$		&	0.22420	&0.22420		\\
$PDF~error~weighted$		&		$\pm$0.00020	&	$\pm$0.00022		\\
(uncertainty in PDF error)&$(~~0.00002)$		&$(~~0.00002)$		\\ 
\hline\hline
\end{tabular}
\label{table_1}
    \end{center}
\end{table}

    \subsubsection{Tevatron pseudo data: default \textsc{nnpdf} 3.0 (\textsc{nnlo}) and default \textsc{nnpdf} 2.3 (\textsc{nnlo}) }  
    

For the  first set of 1600 pseudo experiments  the
 default \textsc{nnpdf} 3.0 (\textsc{nnlo}) is used to generate pseudo data. The  simulated
values of $A_{FB}(M)$  for each experiment  are  compared
 to  $A_{FB}(M)$ templates generated at Tree-level for a range
 of values of $\sin^2\theta_W$ for each of the 
    100 \textsc{nnpdf} 3.0 (\textsc{nnlo}) PDF replicas.  For each replica
 we extract the best fit value of  $\sin^2\theta_W$, the corresponding
statistical error and the fit  $\chi^{2}_{Afb}$.  There are about  500K dimuon events
in each Tevatron pseudo-experiment, which results in a statistical
error  in $\sin^2\theta_W$ of $\pm$0.00042,

An example of the extraction of $\sin^2\theta_W$ from $A_{FB}(M)$  
data at the Tevatron is shown in Fig.\ref{fig_4}. 
 Here,  $\chi^{2}_{Afb}$
is plotted for different values of  $\sin^2\theta_W$ 
The extracted value of
$\sin^2\theta_W$ is the value with the minimum $\chi^{2}_{Afb}$
and the statistical error corresponds to a change of
$\chi^{2}_{Afb}$ by $\pm$1.

For  the second set of 1600 pseudo experiments  the default \textsc{nnpdf} 2.3 (\textsc{nnlo})
is used to generate pseudo data  and the extraction of  $\sin^2\theta_W$ is done using 
  100 \textsc{nnpdf} 2.3 (\textsc{nnlo}) PDF replicas..
   
 For each set, the  extracted value of  $\sin^2\theta_W$  and  the PDF uncertainty 
  are done in two ways. 
    \begin{enumerate}
\item   the standard  average and \textsc{rms}  
of the  $\sin^2\theta_W$ values for the  100 PDF replicas.
 \item  the  $\chi^{2}_{Afb}$ weighted average and weighted \textsc{rms}  of  the 
$\sin^2\theta_W$  values for the  100 PDF replicas.
      \end{enumerate}   
 
\begin{figure}[ht]
\includegraphics[width=8.5 cm, height=6.0 cm]{{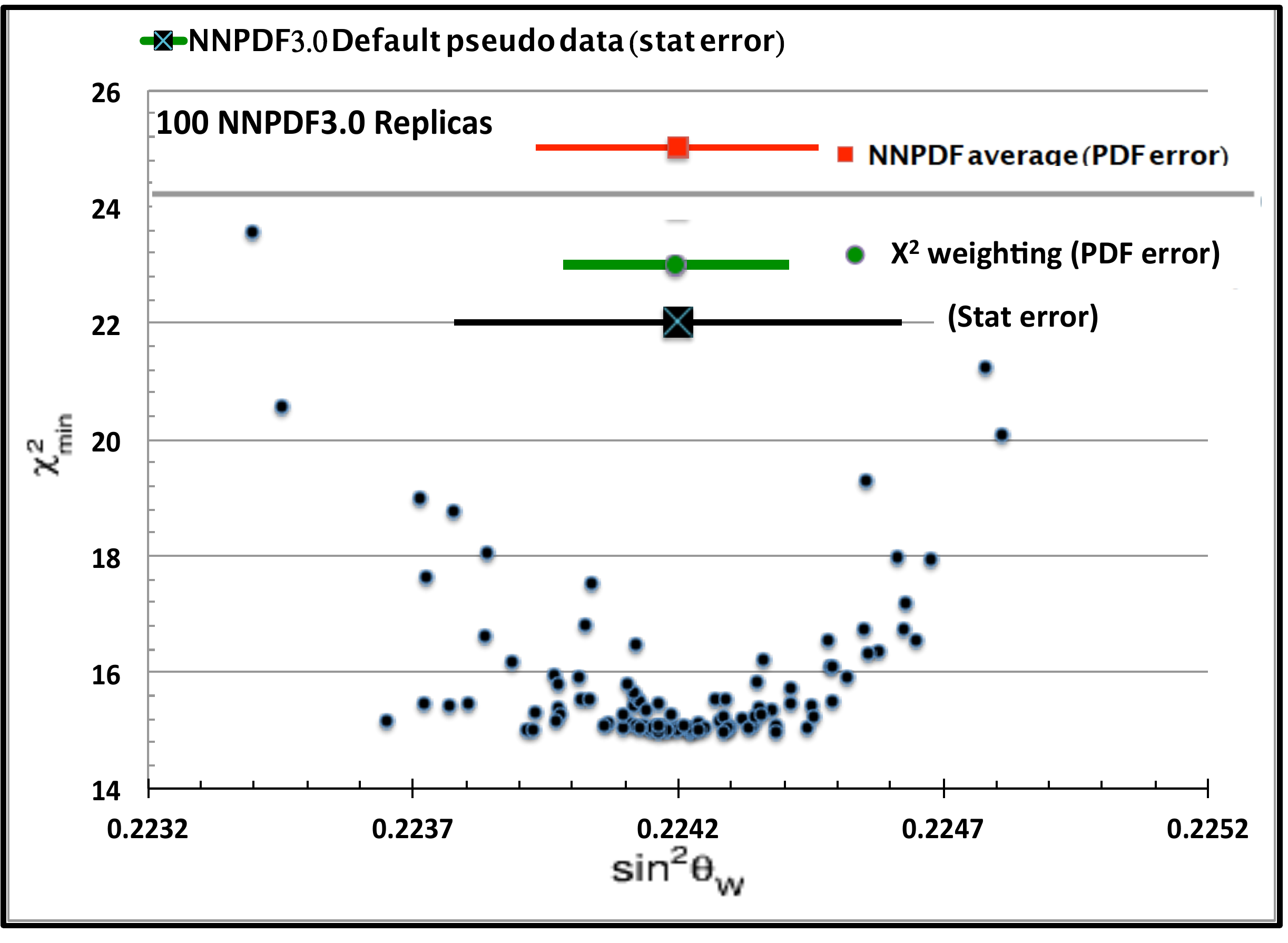}}
\includegraphics[width=8.5 cm, height=6.0 cm]{{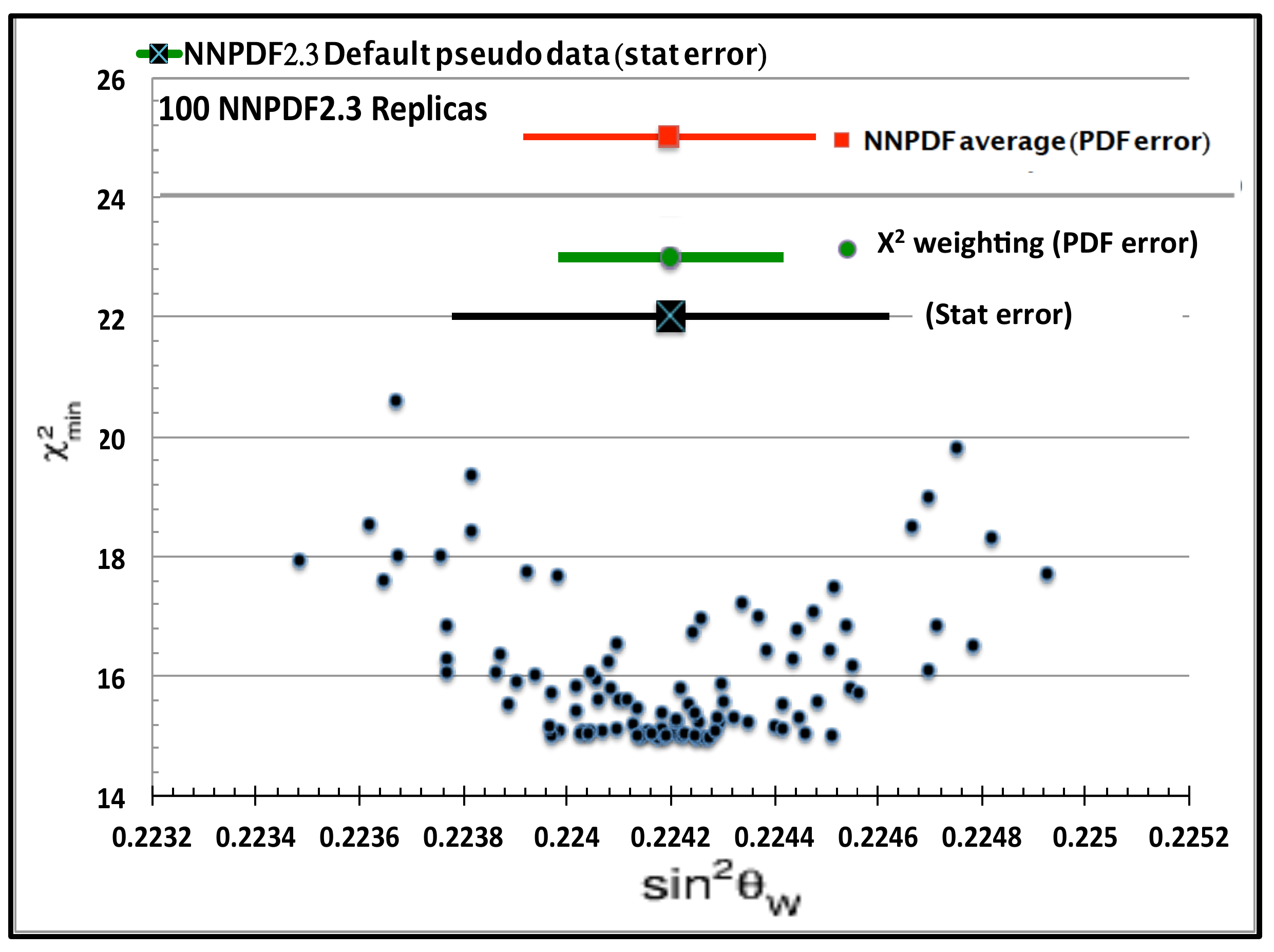}}
\caption{
Tevatron:  A graphical illustration of the
analysis of one  typical pseudo experiment.  Shown is a 
scatter plot of   $\sin^2\theta_W$ and   $\chi^{2}_{Afb}$ values for 100 PDF replicas.
(a) For pseudo experiment generated 
 with the default \textsc{nnpdf} 3.0 (\textsc{nnlo}) and $\sin^2\theta_W$=0.22420
 at Tree-level. 
 (b) or pseudo experiment generated  
 with the default \textsc{nnpdf} 2.3 (\textsc{nnlo})  and $\sin^2\theta_W$=0.22420
 at Tree-level.
 Also shown on the plot is the input value of  $\sin^2\theta_W$
      with the   average statistical error of one  pseudo experiment.
      In addition, we show 
      the  average of the extracted values $\sin^2\theta_W$
      and average  PDF  uncertainty for both  the standard analysis,
      and the  $\chi^{2}_{Afb}$ weighted analysis.
 }
\label{fig_5}
\end{figure}

   For each of  the 100 \textsc{nnpdf} 3.0 (\textsc{nnlo}) (or \textsc{nnpdf} 2.3 (\textsc{nnlo})) replicas we calculate the average of the 1600 extracted
   values of $\sin^2\theta_W$,  the average of the 1600 PDF uncertainties, and the average 1600 statistical errors.  
   These average  quantities have small fluctuation and represent the result 
    of one pseudo experiment on average. 
    The  average
    of the 1600  PDF uncertainties is an estimate of the  typical  uncertainty for one individual pseudo experiment. 
    In order to test for possible bias in the method, the average of the 1600 extracted
   values of $\sin^2\theta_W$ is compared the 0.22420, which is the value used
   in the generation.


    %

    %


    As expected in both analyses the  average extracted
    value of   $\sin^2\theta_W$ is the same as the
    value with which the pseudo data has been generated (0.2242), 
    as shown in  Table~\ref{table_1}.  
    With the  $\chi^{2}_{Afb}$ weighting method the PDF uncertainty in the extracted value of 
     $\sin^2\theta_W$ is reduced from $\pm$0.00027 to $\pm$0.00020.  This illustrates
     that although the statistical error in $\sin^2\theta_W$  of $\pm$0.00042 is somewhat larger than the PDF 
     uncertainty of $\pm$0.00027,
     the $A_{FB}$ data at higher and lower mass  has sufficient precision to constrain the PDFs which
    yields a  25\%   reduction in the PDF uncertainty. 
   
   A graphical illustration of the method  is shown in
    in Fig.~\ref{fig_5} (a) and Fig.~\ref{fig_5} (b).
    For each PDF replica, we calculate the average
    of the extracted values of $\sin^2\theta_W$ and the
      average $\chi^{2}_{Afb}$  of the fits for the 1600 pseudo experiments.
        Fig.~\ref{fig_5} (a) and Fig.~\ref{fig_5} (b) show the scatter
      plot of the average of the extracted values of   $\sin^2\theta_W$  and the
      average  $\chi^{2}_{Afb}$  for the 100 PDF replicas.
      
      Also shown on the plot is the input value of  $\sin^2\theta_W$
      with the   average statistical error of one  pseudo experiment.
      In addition, we show 
      the  average of the extracted values $\sin^2\theta_W$
      and average  PDF uncertainty for both  the standard analysis,
      and the  $\chi^{2}_{Afb}$ weighted analysis.

\begin{table}[htb]
\caption { Tevatron Pseudo data generated
with \textsc{resbos}  and \textsc{cteq} 6.6 PDFs. for a
CDF like detector. Here,
we compare
$\sin^2\theta_W$ values with statistical errors and 
 PDF uncertainties extracted with \textsc{nnpdf} 3.0 (\textsc{nnlo}) PDF replicas
and with   \textsc{nnpdf} 2.3 (\textsc{nnlo})  PDF replicas. 
The values of $\sin^2\theta_W$ extracted
with  \textsc{nnpdf} 3.0 (\textsc{nnlo}) PDF and \textsc{nnpdf} 2.3 (\textsc{nnlo})
are different (for details see text). 
}
    \begin{center}
\begin{tabular}{|c||c|c|c||}
\hline
   Pseudo-Experiment&	\textsc{resbos} 	&	\textsc{resbos} 	\\
Tevatron~10~fb$^{-1}$   & \textsc{cteq}  6.6 &	\textsc{cteq}  6.6   \\
500K~reconstructed  &	 & 	\\
$e^+e^-$ events &	 & 	\\
\hline
$\sin^2\theta_W$ input &	0.22420& 0.22420 	\\
\hline\hline
statistical~error 	&	$\pm$0.00042	&	$\pm$0.00042	\\
$\Delta\sin^2\theta_W$		&		&		\\
CT10~PDF~error& $\pm$0.00026 & $\pm$ 0.00026\\
\hline \hline
Number of analysis replicas	&100		&	100	\\ 
\textsc{nnpdf}  replica set 	&\textsc{nnpdf} 3.0 	&\textsc{nnpdf} 2.3 		\\ 
Templates		&	Tree-level 	&Tree-level		\\
		&		&		\\ 
		Average  method &	$N_{eff}=100$			&$N_{eff}=100$	\\
extracted ~$\sin^2\theta_W$		&	0.22425	&0.22469		\\ 
bias& +0.00005& +0.00049\\
PDF~error~\textsc{rms}    	&		$\pm$0.00027	&	$\pm$0.00027		\\
(uncertainty in PDF error)&$(~~0.00002)$		&$(~~0.00002)$		\\ 
&		&		\\ 
{$A_{FB}$ $\chi^{2}_{Afb}$ weighting}		&	$N_{eff}=88$			&$N_{eff}=63$		\\
extracted ~$\sin^2\theta_W$		&	0.22425	&0.22452		\\
bias& +0.00005& +0.00032\\
PDF~error~weighted		&		$\pm$0.00020	&	$\pm$0.00021\\
(uncertainty in PDF error)&$(~~0.00002)$		&$(~~0.00002)$		\\
\hline\hline
\end{tabular}
\label{table_2}
    \end{center}
\end{table}

\begin{figure}[ht]
\includegraphics[width=8.45 cm, height=6.0 cm]{{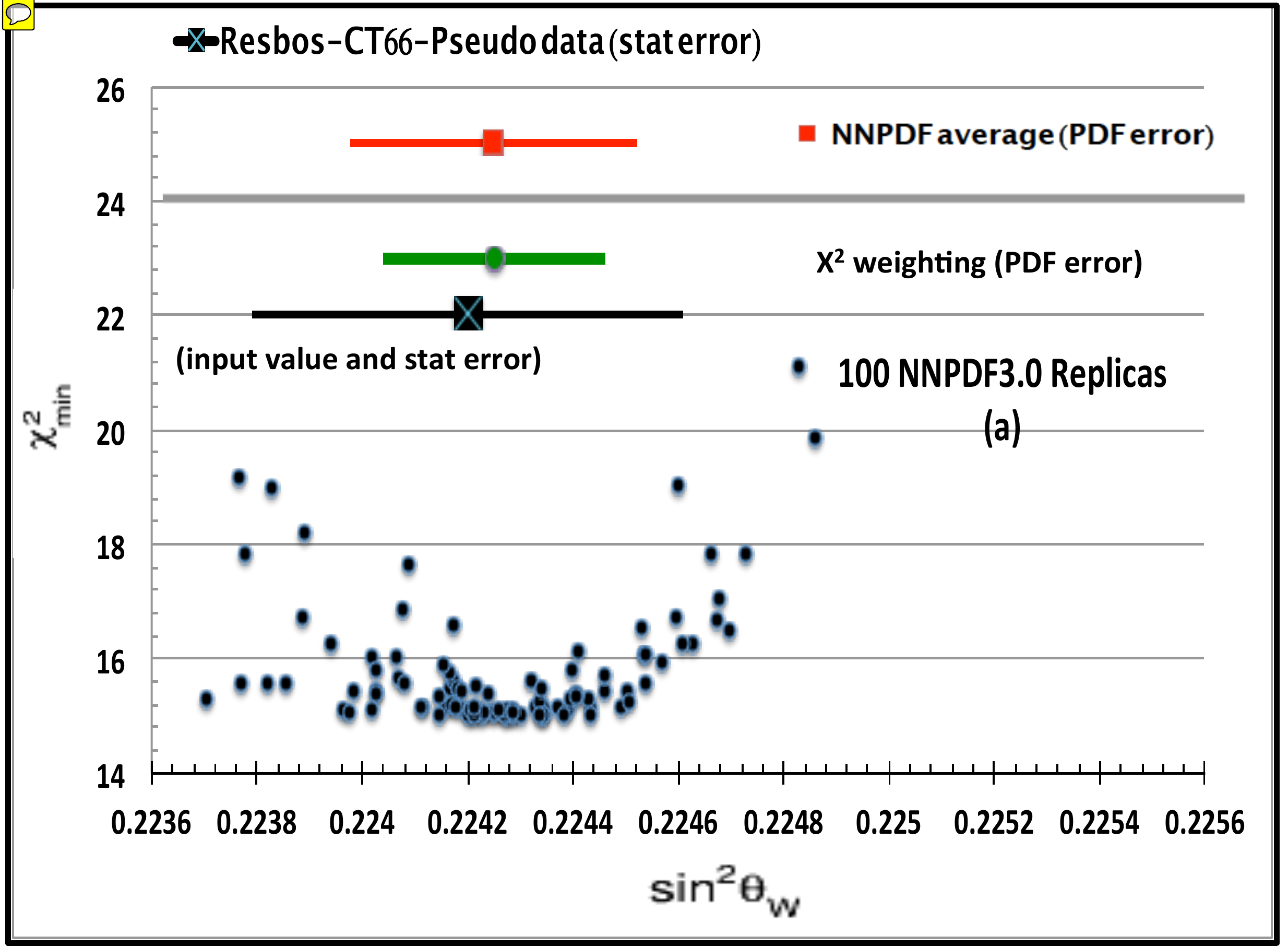}}
\includegraphics[width=8.5 cm, height=6.0 cm]{{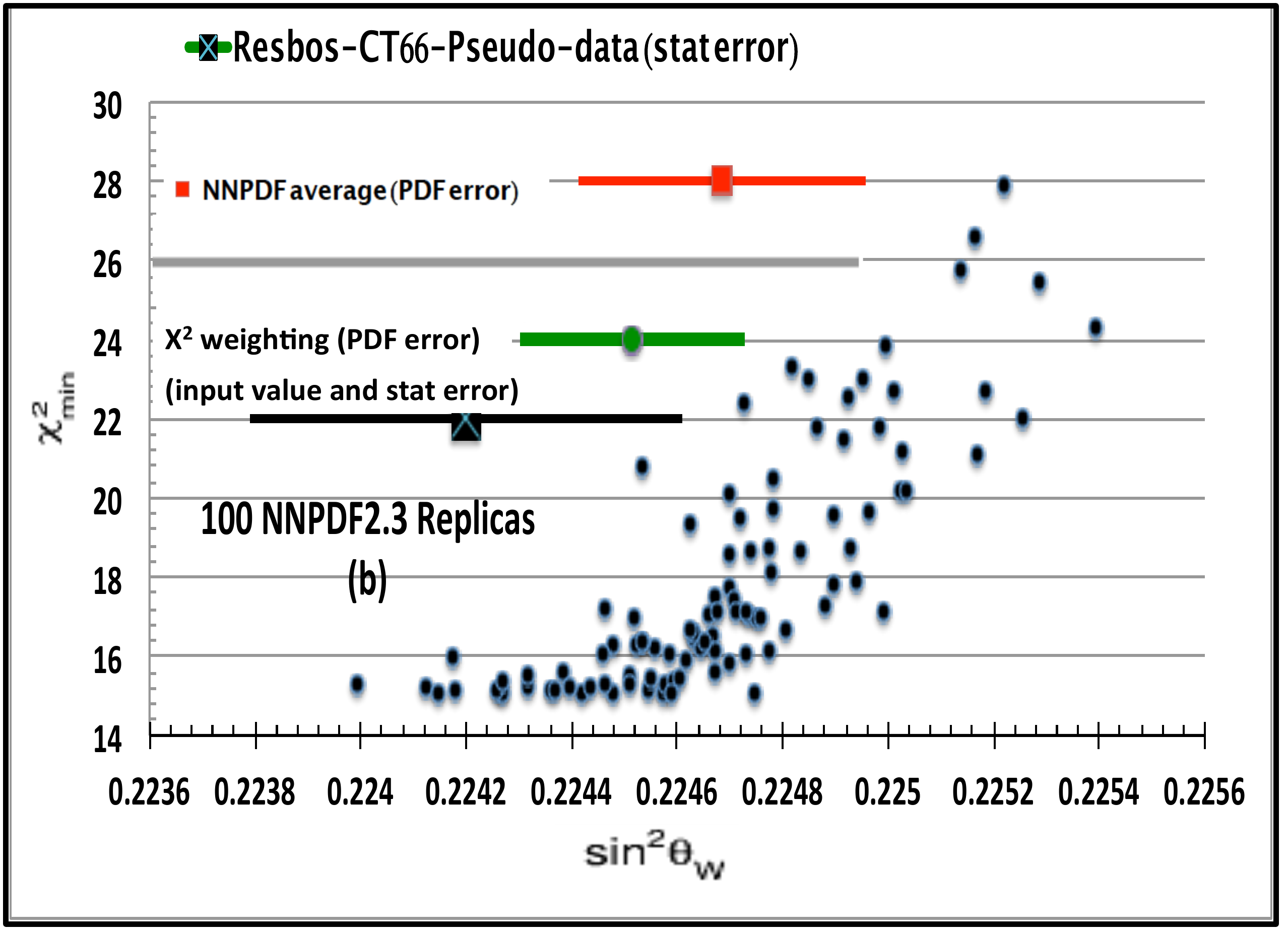}}
\caption{ Analysis of a Tevatron pseudo-experiment.
The  pseudo data are generated 
by \textsc{resbos}  with  \textsc{cteq}  6.6 PDF and $\sin^2\theta_W$=0.22420.
This figure  illustrates that  with the $\chi^{2}_{Afb}$ weighting method
 we can determine that pseudo data 
generated with \textsc{cteq}  6.6 PDFs are not  consistent  
with the \textsc{nnpdf} 2.3 (\textsc{nnlo}) set.
 (a) Analysis with  100 \textsc{nnpdf} 3.0 (\textsc{nnlo}) replicas.
 (b) Analysis with 100 \textsc{nnpdf} 2.3 (\textsc{nnlo}) replicas.
 The distribution
                 of $\chi^{2}_{Afb}$ values versus $\sin^2\theta_W$  provides
                   a powerful tool to discriminate against PDF sets
                     which are incompatible with the data.  The PDF sets
                       which are compatible with the data should have a symmetric
                       distribution of $\chi^{2}_{Afb}$ values versus $\sin^2\theta_W$.
                       }
\label{fig_6}
\end{figure}

 \subsubsection{Pseudo data: \textsc{resbos}  with \textsc{cteq}  6.6 PDF set } 
 
 We perform two analyses of  the third set of 1600 pseudo experiments  (\textsc{cteq} 6.6 pseudo data).
In one analysis the  simulated values of $A_{FB}(M)$  for each experiment are  compared
 to  templates calculated  at  Tree-level for each of the  100 \textsc{nnpdf} 3.0 (\textsc{nnlo}) PDF replicas.
 In the other analysis the  simulated values of $A_{FB}(M)$  for each experiment are  compared
 to  templates calculated at Tree-level for each of the  100 \textsc{nnpdf} 2.3 (\textsc{nnlo}) PDF replicas.
   %
    In each of the two analyses,    $\sin^2\theta_W$ is
     extracted using both   the standard  average and \textsc{rms} , and  also the $\chi^{2}_{Afb}$ weighted average 
    and \textsc{rms}  of the 100 PDF replicas. 
    The results are summarized in Table \ref{table_2}.

     In the analysis of the   \textsc{resbos} /\textsc{cteq} 6.6  pseudo data
      with \textsc{nnpdf} 3.0 (\textsc{nnlo}) replica templates
          we find   that  the   PDF uncertainty in the extracted  value of 
     $\sin^2\theta_W$ when we use the standard average is $\pm$0.00027.
     The PDF uncertainty is  reduced to $\pm$0.00020 when
     the $\chi^{2}_{Afb}$ weighting method is used, 
     as shown in Table~\ref{table_2} and Fig.~\ref{fig_6}.
     The effective number of replicas is reduced from 100 to 88.
     The  average  value is   $\sin^2\theta_W=0.22425$  
     for both the standard analysis  and the  $\chi^{2}_{Afb}$ 
     weighting analysis. The very small difference (+0.00005)
     from the input value of  $\sin^2\theta_W=0.22420$  is attributed
     to the difference between the  \textsc{resbos}  pseudo data which
     is generated at  \textsc{nlo}  and the templates which
     were done at   LO Tree-level.

      In contrast, the standard analysis  with  the \textsc{nnpdf} 2.3 (\textsc{nnlo}) replica
      templates yields  a value which is biased by +0.00049$\pm$0.00001. 
      This is larger than the PDF uncertainty of  $\pm$0.00027.  
      This  bias indicates that the \textsc{nnpdf} 2.3 (\textsc{nnlo})
      set is not fully consistent with the \textsc{cteq} 6.6 PDF for the Bjorken $x$
      region for the production of Z bosons at the Tevatron. 
       When the $\chi^{2}_{Afb}$ weighting technique is used instead, 
       the bias is  partially reduced from  +0.00049$\pm$0.00001 to +0.00032$\pm$0.00001,
        and the  effective number of PDFs is reduced from 100 to 63. 
         The reduced bias is expected  because $\chi^{2}_{Afb}$ weighting assigns
           small weights to a fraction of   \textsc{nnpdf} 2.3 (\textsc{nnlo})   PDF replicas which are
            incompatible with the \textsc{cteq} 6.6. pseudo data.

                As shown in Fig.~\ref{fig_6} the distribution
                 of $\chi^{2}_{Afb}$ values versus $\sin^2\theta_W$  provides
                   a powerful tool to discriminate against PDF sets
                     which are incompatible with  each other or with the data. 
                     Our study indicates that  \textsc{cteq}  6.6 PDFs are
                     inconsistent with the  \textsc{nnpdf} 2.3 (\textsc{nnlo}) set, but are  consistent
                      with the \textsc{nnpdf} 3.0 (\textsc{nnlo}) set.  One of the difference between
                          \textsc{nnpdf} 3.0 and \textsc{nnpdf} 2.3  is that \textsc{nnpdf} 2.3 used $W$ asymmetry
                          data which is  now known to be incorrect. 
                          %
 \begin{figure}[ht]
\includegraphics[width=8.5 cm, height=9.0 cm]{{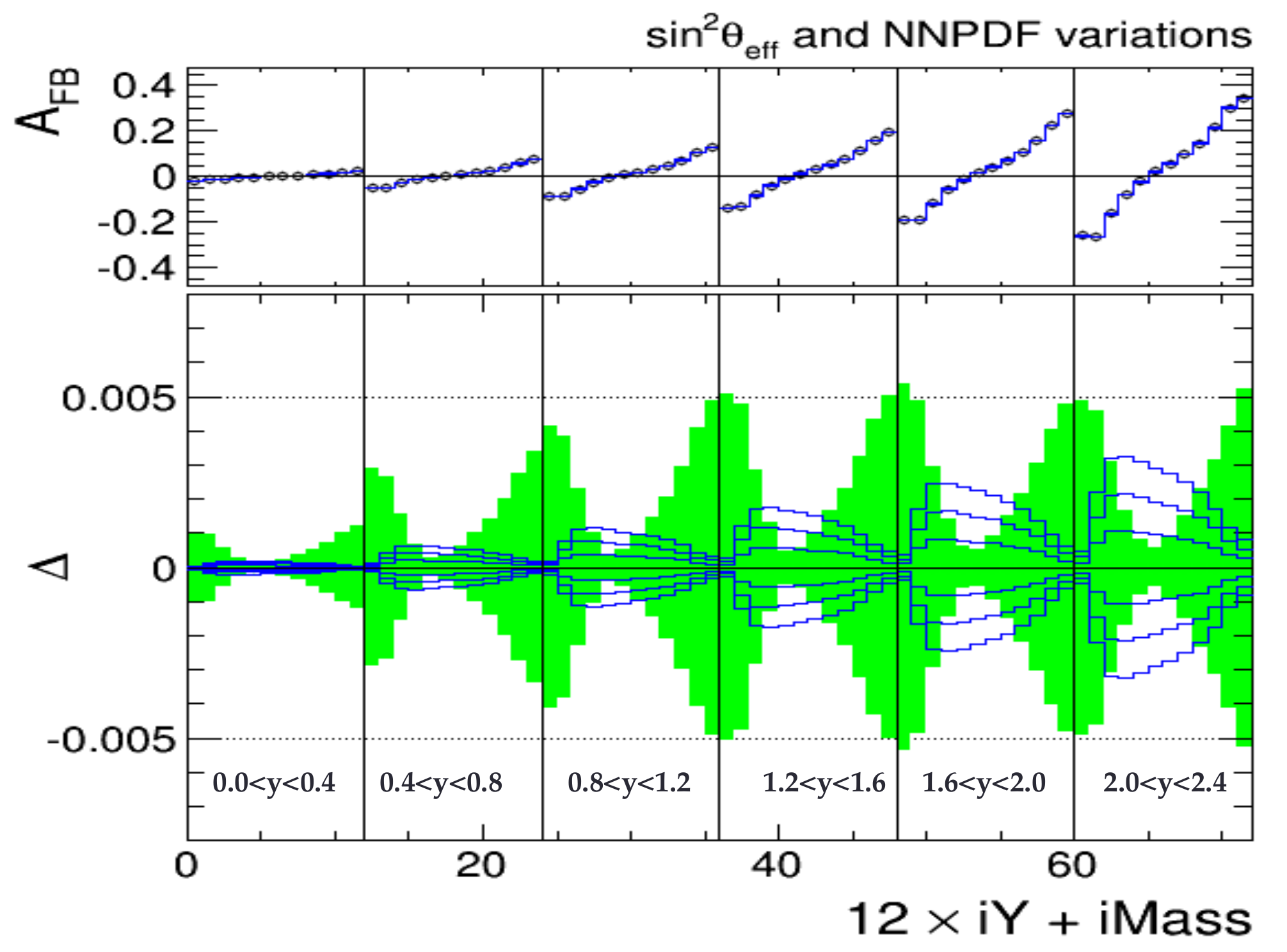}}
\caption{
LHC:  Top panel:  $A_{FB}$ at the LHC at $\sqrt{s}$=8 TeV for
six rapidity bins  (iY=0 to 5) with  average $|y|$ values of  0.2, 0.6, 1.0, 1.4, 1.8 and 2.2.
For each rapidity bin there are twelve mass bins discussed in the text (iMass=0 to 11).
 The horizontal scale for each of  the six plots is the dimuon invariant mass
 for each rapidity bin expressed as $12\times iY + iMass$.
Bottom panel:  The green bands span the  difference between $A_{FB}(M)$ 
 calculated for the 100 \textsc{nnpdf} 3.0 (\textsc{nlo}) replicas and $A_{FB}(M)$ calculated
for  the central  default \textsc{nnpdf} 3.0 (\textsc{nlo})   for the six dimuon rapidity bins. The blue lines are
the  differences between 
$A_{FB}(M)$  calculated with  different values of $\sin^2\theta_{eff}$ (0.23120 $\pm$0.00040, $\pm$0.00080 and $\pm$0.00120).
and the values calculated with nominal  $\sin^2\theta_{eff}$=0.23120.
For all of the blue lines,  $A_{FB}(M)$ is  calculated
with  the central  default \textsc{nnpdf} 3.0 (\textsc{nlo}). The calculations
are done with the \textsc{powheg}  MC generator.}
\label{fig_7}
\end{figure}
\section {Production of dilepton events at the LHC}
At the LHC, dileptons are produced by annihilation
of quarks in one proton with antiquarks in the
other proton.
\begin{eqnarray}
\frac{d \sigma}{dMdy}(pp) }\propto  { \sum_{flavor}
{ v_{i}\{ {q}_i (x_{1})  {\bar q} _i (x_{2}) +{\bar q}_i (x_{1})  q_i (x_{2})\} 
}
\label{siglhc}
\end{eqnarray}
Because on average, quarks carry
more momentum than antiquarks, the quark direction
is assumed to be the direction of motion 
of the dilepton pair.   This is more likely to be
true for dileptons produced at high rapidity.
At the LHC the asymmetry from the first term
of equation \ref{siglhc} is diluted by
the asymmetry of the second term (which is in the
 opposite direction). 
Equation \ref{siglhc} shows that for
y=0 ( $x_1=x_2$)  the asymmetries
for the two terms cancel each other.  

An estimate of the dilution
of  $A_{FB}(M)$  can be obtained  from the 
 probability to misidentify the direction of the quark
$f(M,y)$. For $pp$ collisions $f(M,y)$ is the
fraction of events for which the antiquark carries
more momentum than the quark.

\begin{equation}
f(M,y) \approx \frac{ \sum_{flavor} { v_{i} \{
{\bar {q}_i(x_{1}})~{{q}_i(x_{2})}} \} }
 { \sum_{flavor}{ v_{i}
\{ {q}_i(x_{1})   {\bar q}_i (x_{2}) +{\bar q}_i(x_{1}) q_i (x_{2})\} }} 
\end{equation}

The asymmetry is significant only when $x_1$ is large and $x_2$
is small (when  $x_2$ is small,  
$u(x_2) \approx d(x_2) \approx \bar{u} (x_2) \approx \bar{d}(x_2)$).
The  asymmetry for $u$ quarks dominates, and the fractions
of $d$ quarks and $\bar u$ antiquarks are sources
of dilution.
\begin{eqnarray}
\label{eq-lhc}
D ^{LHC}_{AFB}(d)  &\appropto&  \frac {d(x_1)\bar{d}(x_2)}{u(x_1)\bar{u}(x_2)} \approx   \frac{d}{u} (x_1)\\
D ^{LHC}_{AFB}(\bar{q}) &\appropto& \frac {\bar{u}(x_1) u(x_2)}   {u(x_1) \bar{u(}x_2)} \approx \frac{\bar{u}}{u}(x_1)
\end{eqnarray} 

Since $x_{1} = \frac{M}{\sqrt{s} }e^{+ y}$ both the mass and rapidity dependence
of $A_{FB}$ provides information on PDFs.

 %
At the LHC, the W asymmetry also provides information
on the d/u ratio. The $W^-/W^+$ ratio  at the LHC can be written as
  \begin{eqnarray}
(\frac{W^-}{W^+})^{LHC}&\approx & \frac {d( x_1) ~\bar u(x_2) +s( x_1) ~\bar c(x_2)  } {u (x_1) ~ \bar d (x_2) +c( x_1) ~\bar s(x_2)} \\ \nonumber
&\approx& \frac {d/u(x_1)} {\bar d/ \bar u (x_2)} \approx \frac{d}{u} (x_1)
\label{sig}
\end{eqnarray}

%
Unlike the situation at the Tevatron, more precise
W asymmetry measurements at the LHC provide information on  the
absolute value of  $\frac{d}{u} (x_1)$. Therefore, new
measurements of the W charge asymmetry at the LHC (which have
not yet been incorporated into PDF fits) can be used
in combination with  the constraints from $A_{FB}$
to reduce the PDF uncertainty in the
extractions of  $\sin^2\theta_{eff}$  and
$\sin^2\theta_W$ at the LHC.

Combining constraints from both  $A_{FB}$ and
new  W asymmetry measurements can be done by
adding the values of  $\chi^2_{Wasym}$ from the comparison 
of the new W asymmetry data with the predicted
W asymmetry for each PDF replica, to the $\chi^{2}_{Afb}$ values
from the fits to extract  $\sin^2\theta_{eff}$ from
the  $A_{FB}(M,y)$ data for  each PDF replica.

\subsection{Mass dependance of $A_{FB}(M,y)$ as a function of  $\sin^2\theta_{eff}$ and PDFs at the LHC}

The top panel of Fig. \ref{fig_7}  shows $A_{FB}(M,y)$ at the LHC at $\sqrt{s}$=8 TeV for
six rapidity bins   $0<|y|<0.4$, $0.4<|y|<0.8$, $0.8<|y|<1.2$, $1.2<|y|<1.6$, $1.6<|y|<1.0$ and $2.0<|y|<2.4$ (iY=0 to 5). 
These six bins have  average $|y|$ values of  0.2, 0.6, 1.0, 1.4, 1.8 and 2.2.
The mass bins are 60-70,   70-78,   78-84, 84-87, 87-89, 89-91, 91-93, 93-95,  95-98,  98-104,  104-112 and 112-120 GeV.
 The horizontal scale for each of  the six plots is the dimuon invariant mass
 for each rapidity bin expressed as $12\times iY + iMass$.

The calculations are done with the \textsc{powheg} \cite{POWHEG}  MC generator. The
version of \textsc{powheg}   that is used does not include electroweak radiative corrections.
Therefore, this version of  \textsc{powheg}  requires an input value of  $\sin^2\theta_{eff}$ for the  calculation of $A_{FB}$

 
The green bands in the bottom panel of  Fig. \ref{fig_7} span  the  difference between $A_{FB}(M,y)$ 
calculated with the 100 \textsc{nnpdf} 3.0 (\textsc{nlo}) replicas and $A_{FB}(M,y)$ calculated
with the  default \textsc{nnpdf} 3.0 (\textsc{nlo}) PDF. 

 The blue lines are the  differences between 
$A_{FB}(M,y)$  calculated for several values of  $\sin^2\theta_{eff}$
($\sin^2\theta_{eff}$=0.23120 $\pm$0.00040, $\pm$0.00080 and $\pm$0.00120) 
and $A_{FB}(M,y)$ for the nominal  $\sin^2\theta_{eff}$=0.23120.
 For all of the blue lines,  $A_{FB}(M,y)$ is  calculated
with  the default  \textsc{nnpdf} 3.0(\textsc{nlo})  PDF.
\begin{figure}[ht]
\includegraphics[width=8.5 cm, height=6.0 cm]{{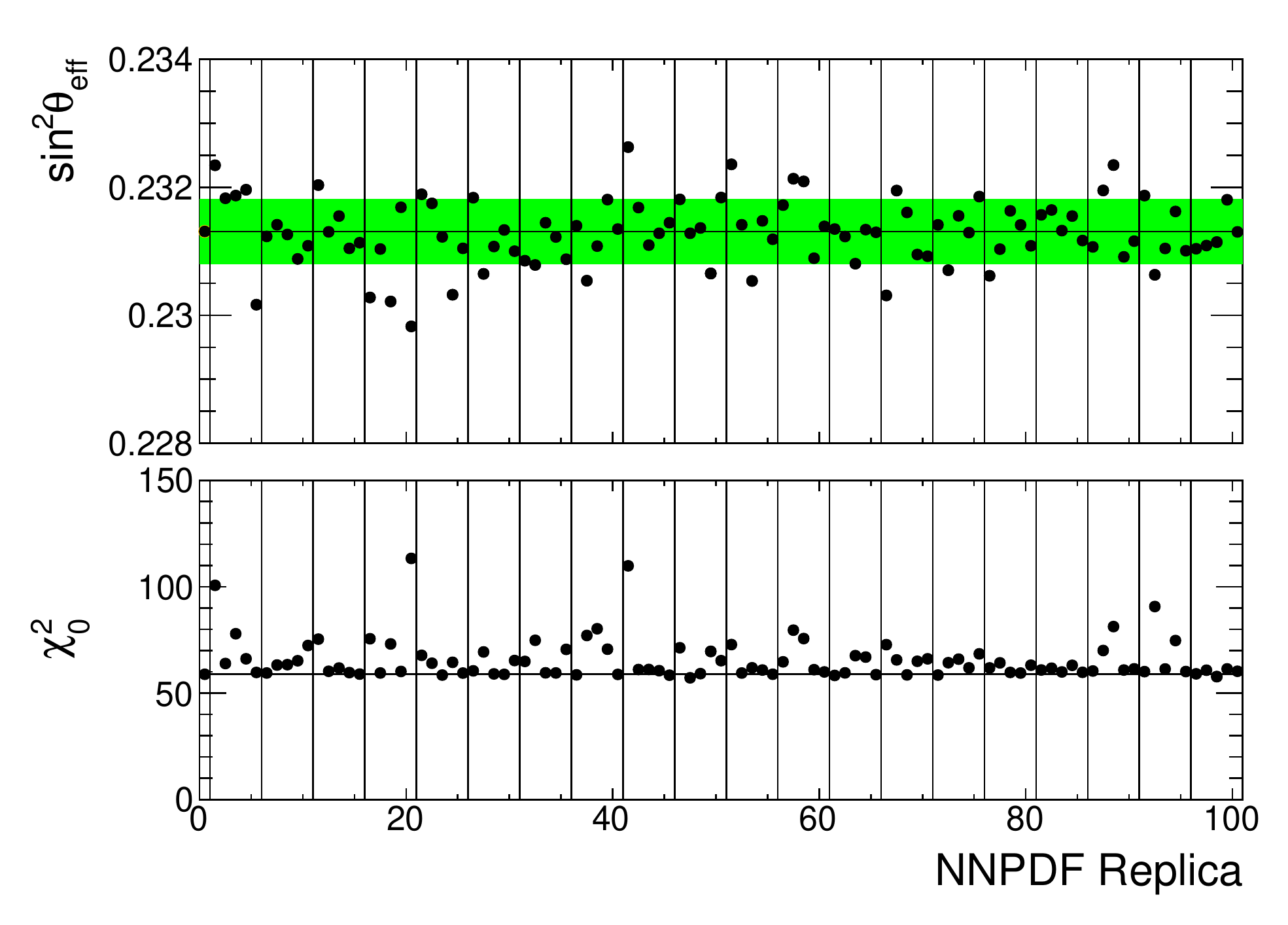}}
\includegraphics[width=8.7 cm, height=4.0 cm]{{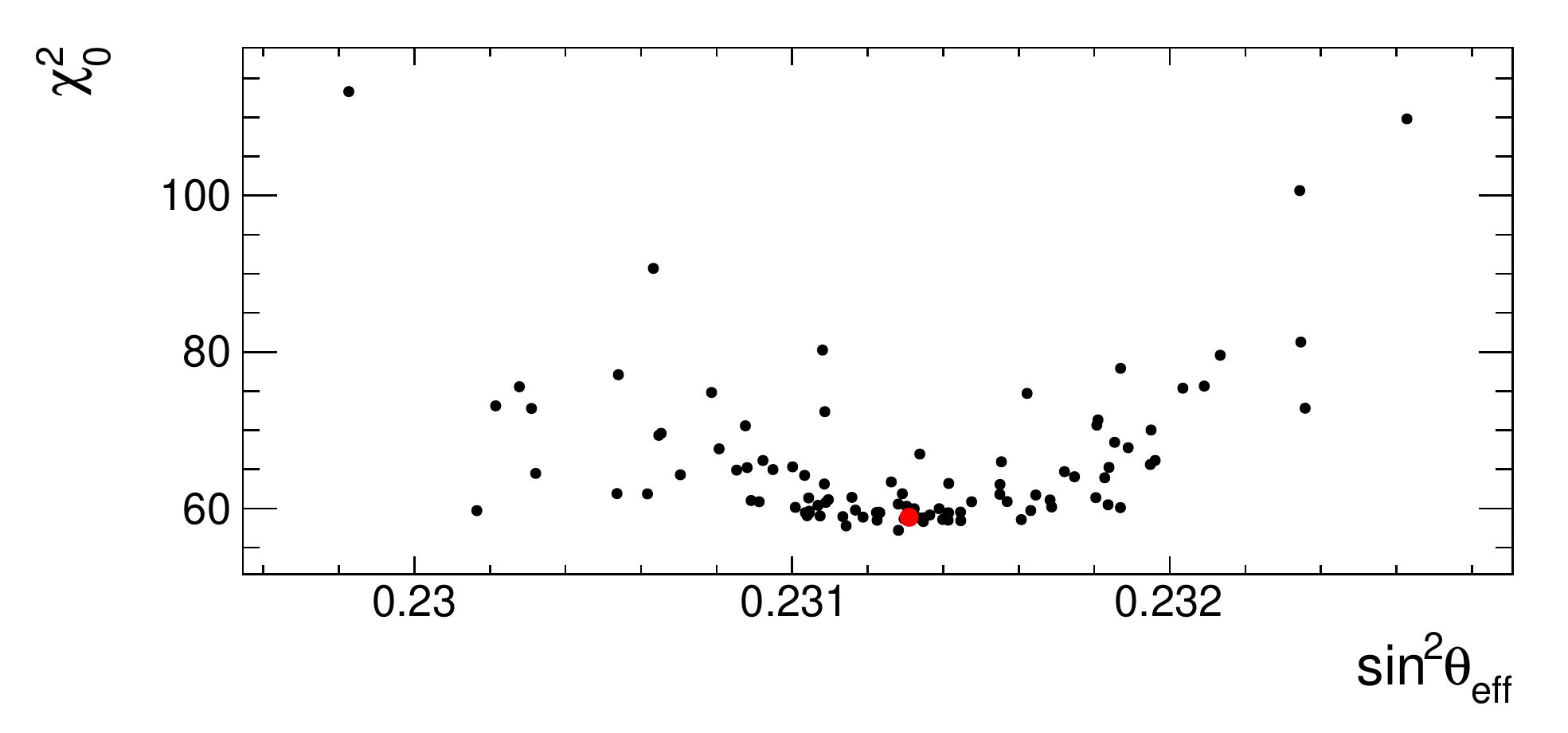}}
\caption{  Analysis of one of the 64  LHC pseudo experiments
 (6.7 M  dimuon events with  CMS-like detector acceptance cuts) with 100 PDF replicas.
 The pseudo data are  generated 
by the \textsc{powheg}  MC with  the default  \textsc{nnpdf} 3.0 (\textsc{nlo}) PDF and $\sin^2\theta_{eff}$=0.23120.
 The top two panels show the  extracted $\sin^2\theta_{eff}$ and 
  corresponding  $\chi^{2}_{Afb}$  values  from fits
 to $A_{FB}(M,y)$ versus  replica
 number for  the 100 \textsc{nnpdf} 3.0 (\textsc{nlo}) replicas.
 The bottom panel shows the same results
 in the form of a scatter plot of $\chi^{2}_{Afb}$ values versus $\sin^2\theta_{eff}$
 for one pseudo experiment.  The number of degrees of freedom is 71
 (=$6\times 12-1$).
  } 
 \label{fig_8}
\end{figure}

As is the case for the Tevatron, the dependence of  $A_{FB}(M,y)$ on $\sin^2\theta_{eff}$
 and  on PDFs is different.    In the region of the $Z$ pole,   $A_{FB}(M,y)$ 
 is sensitive to the vector couplings, which are  functions of    $\sin^2\theta_{eff}$.
 At higher and lower mass  $A_{FB}(M,y)$ is sensitive to the
 axial coupling and therefore insensitive to value of  $\sin^2\theta_{eff}$.
 As is the case for the Tevatron, the magnitude of the  dilution of  $A_{FB}(M)$
  is larger in regions where the absolute value of   $A_{FB}(M)$ is large
  (i.e. away from the $Z$ pole). At the  LHC 
  the  dilution depends   on  both M and y.
 The  combined mass and rapidity dependence 
  of the dilution at the LHC provides  more stringent 
 constraints on PDFs  than $A_{FB}(M)$ measurements at the Tevatron.
     
     \subsection {MC studies with 
     NNPDF  3.0 PDFs at the LHC}

For studies of  $A_{FB}(M,y)$ at the LHC we 
simulate Drell-Yan  dimuon data for 64  pseudo experiments
for a  CMS like detector at $\sqrt{s}$=8 TeV.
The pseudo data are  generated by
 the  \textsc{powheg}   \textsc{nlo}  MC
generator with the  default  \textsc{nnpdf} 3.0 (\textsc{nlo}) PDFs, The pseudo data are generated 
with an effective mixing angle  $\sin^2\theta_{eff}$=0.23120.

 For each pseudo experiment,
 we generate a sample of 15.6  Million dimuon events with $M_{\mu\mu} >$ ~50 GeV, which  corresponds to an integrated
luminosity of 15.0 fb$^{-1}$. 
 This is similar  to the  $\approx$19 fb$^{-1}$ of  integrated luminosity
collected by CMS and ATLAS  at 8 TeV.  We apply acceptance and transverse
momentum cuts which are similar to a CMS-like detector. We also smear the muon energy with
a muon momentum resolution similar to a CMS-like detector. The final sample consists
 6.7M reconstructed dimuon events.

The 8 TeV $W$  decay lepton asymmetry data at the LHC has not yet been incorporated
into the most recent PDF fits.  Therefore, 
in addition to  $A_{FB}(M,y)$, we also use the default \textsc{nnpdf} 3.0 (\textsc{nlo}) and  generate pseudo data
for the  W muon decay asymmetry as a function of muon rapidity (for muon
transverse momentum PT$>$25 GeV). This simulates the  W asymmetry 
measurement at 8 TeV.

In the analysis of each of the   64 pseudo experiments  generated
with the default \textsc{nnpdf} 3.0 (\textsc{nlo})  the  extracted 
values of $A_{FB}(M,y)$  for each experiment are  compared
 to  $A_{FB}(M,y)$ templates. The templates are generated with the \textsc{powheg}  MC for a range
 of values of $\sin^2\theta_{eff}$ for each of the 
    100 \textsc{nnpdf} 3.0 (\textsc{nlo}) PDF replica.  For each replica
 we extract the best fit value of  $\sin^2\theta_{eff}$, the corresponding
statistical error and the fit  $\chi^{2}_{Afb}$.  

In addition, we calculate $\chi^2_{Wasym}$ which is the $\chi^2$
for the agreement between the predictions for the W lepton decay
asymmetry and the W lepton decay asymmetry pseudo data at 8 TeV
for each  of the 100 PDF replicas.


Fig. \ref{fig_8}  shows the results from one of the 
64 pseudo experiments at the LHC.  The top two panels 
show the  extracted $\sin^2\theta_{eff}$ and 
  corresponding  $\chi^{2}_{Afb}$  values  from fits
 to $A_{FB}(M,y)$ versus  replica
 number for  the 100 \textsc{nnpdf} 3.0 (\textsc{nlo}) replicas.
 The bottom  shows  the same results in the form of a 
   scatter plot of $\chi^{2}_{Afb}$ values versus $\sin^2\theta_{eff}$
   for one pseudo experiment.  The number of degrees of freedom is 71
 (=$6\times 12-1$).
   %
 \begin{figure}[ht]
\includegraphics[width=8.5 cm, height=4 cm]{{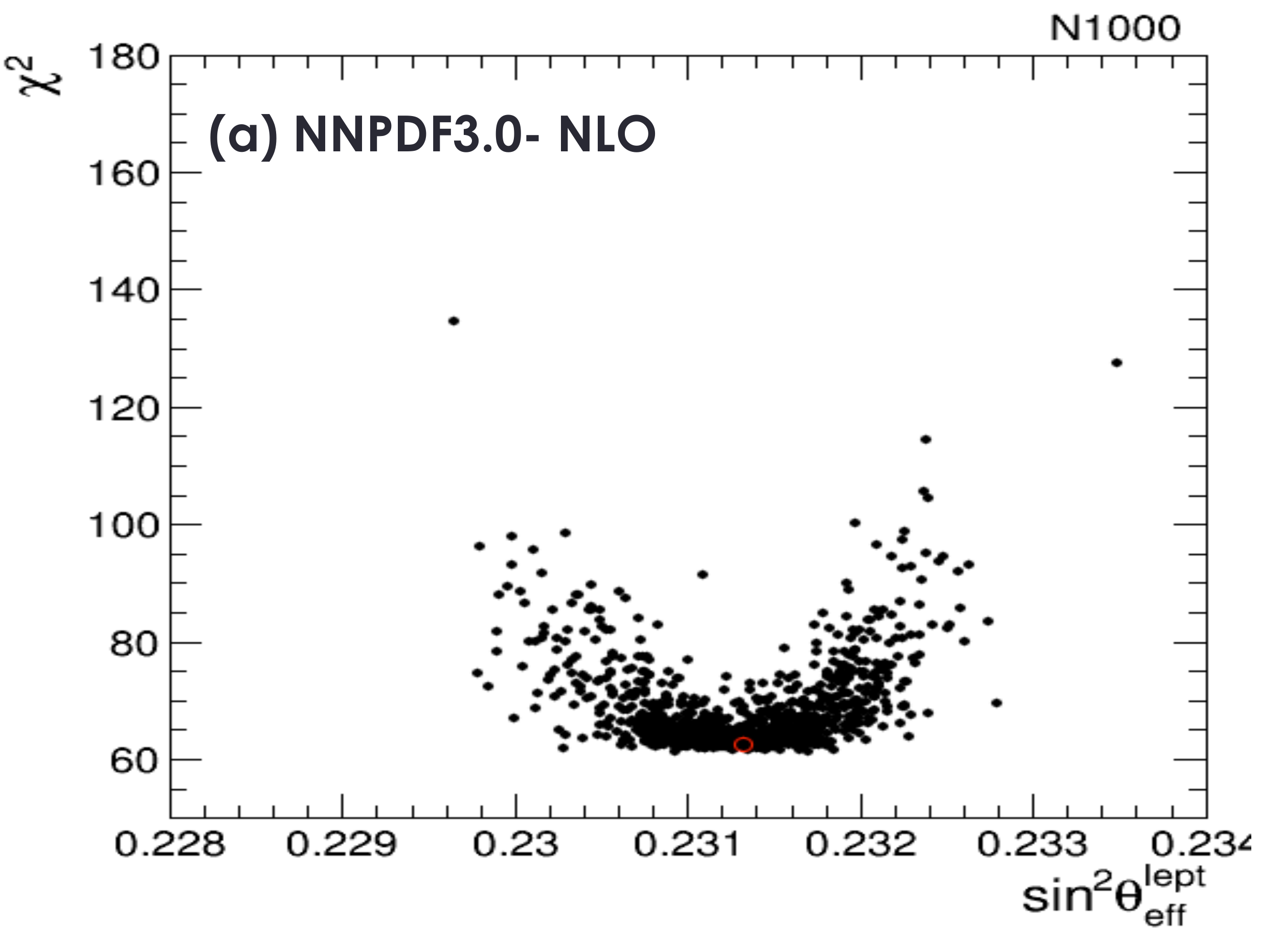}}
\includegraphics[width=8.55 cm, height=4 cm]{{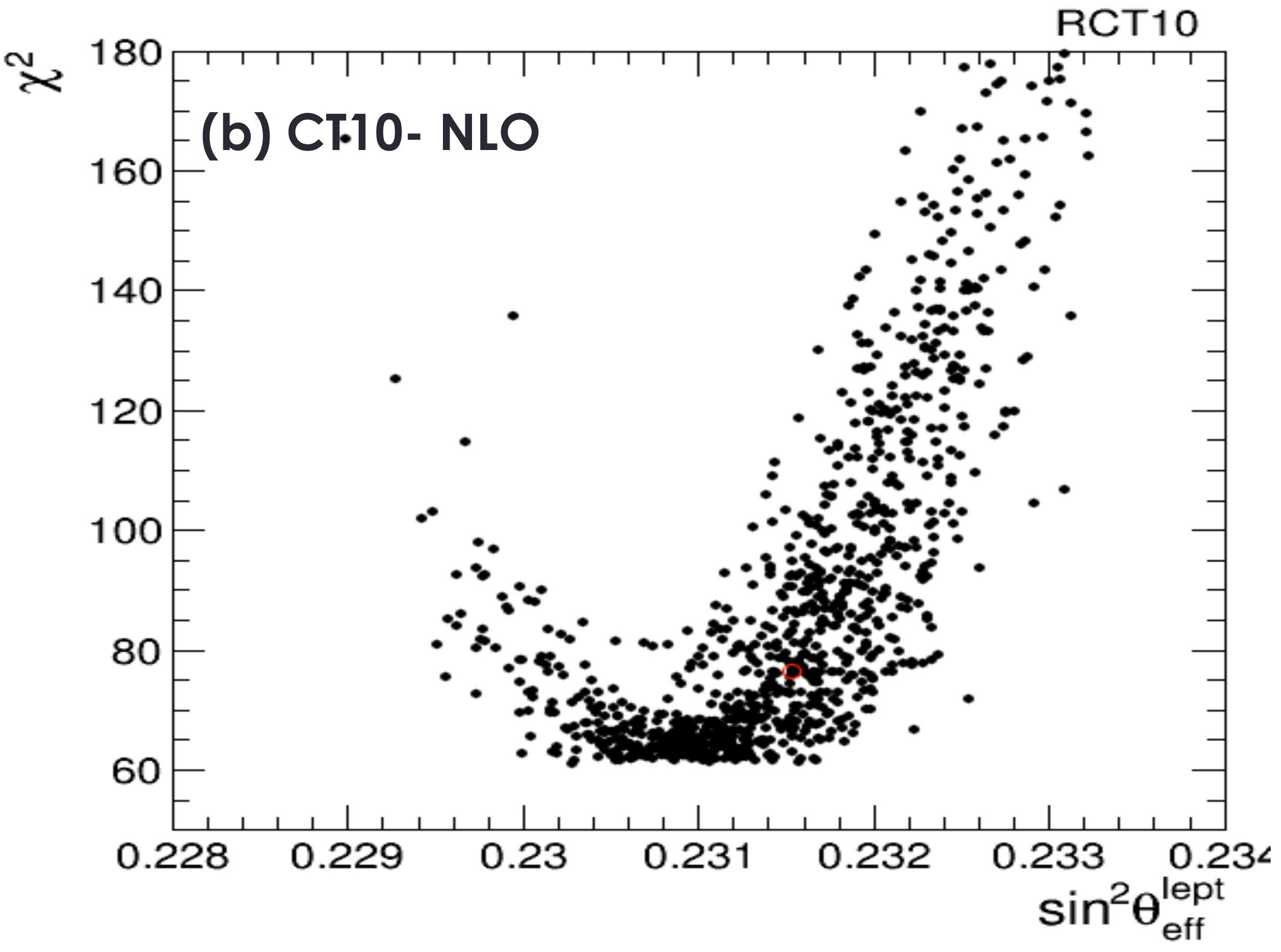}}
\includegraphics[width=8.4 cm, height=4 cm]{{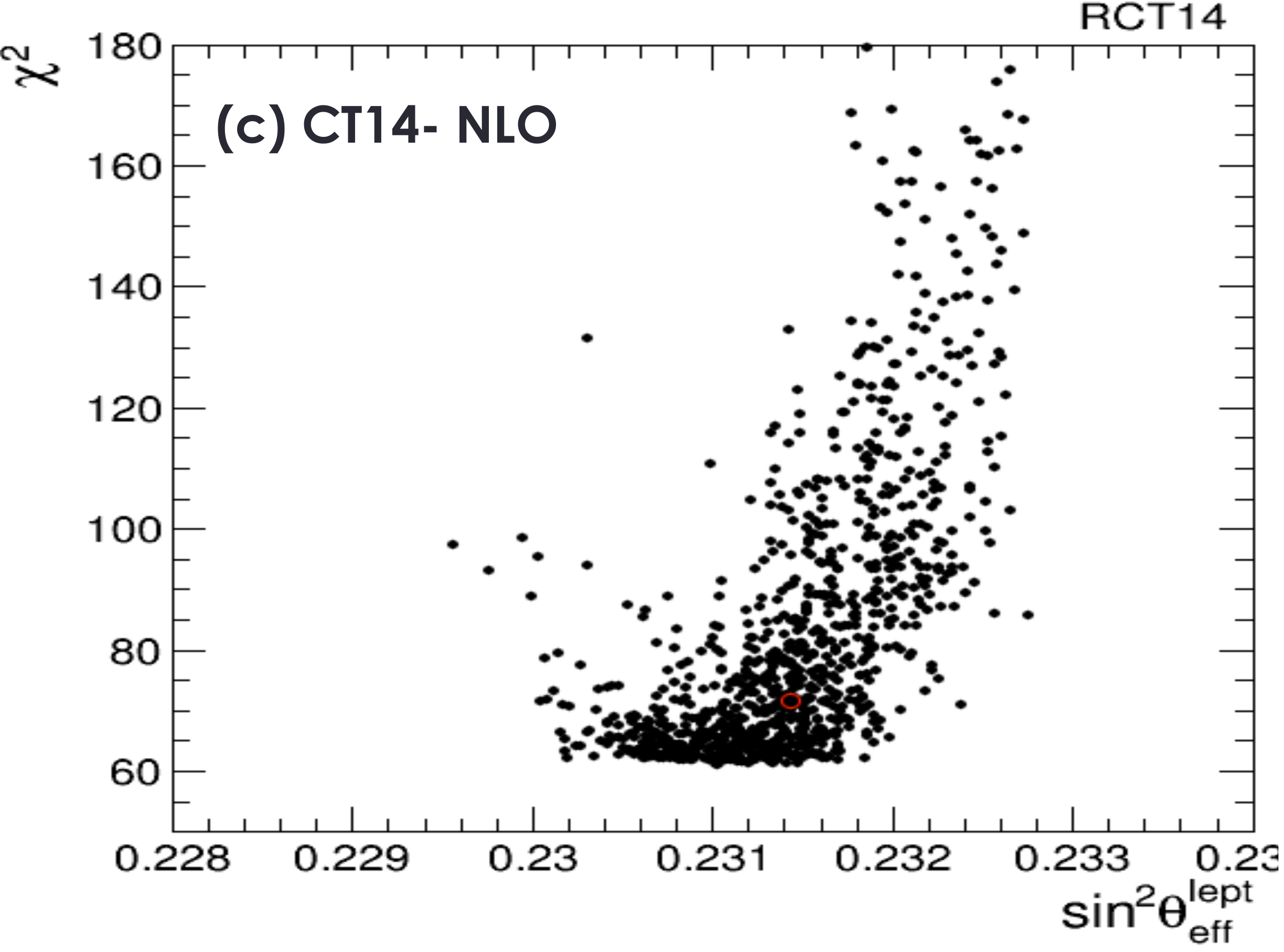}}
\includegraphics[width=8.55 cm, height=4  cm]{{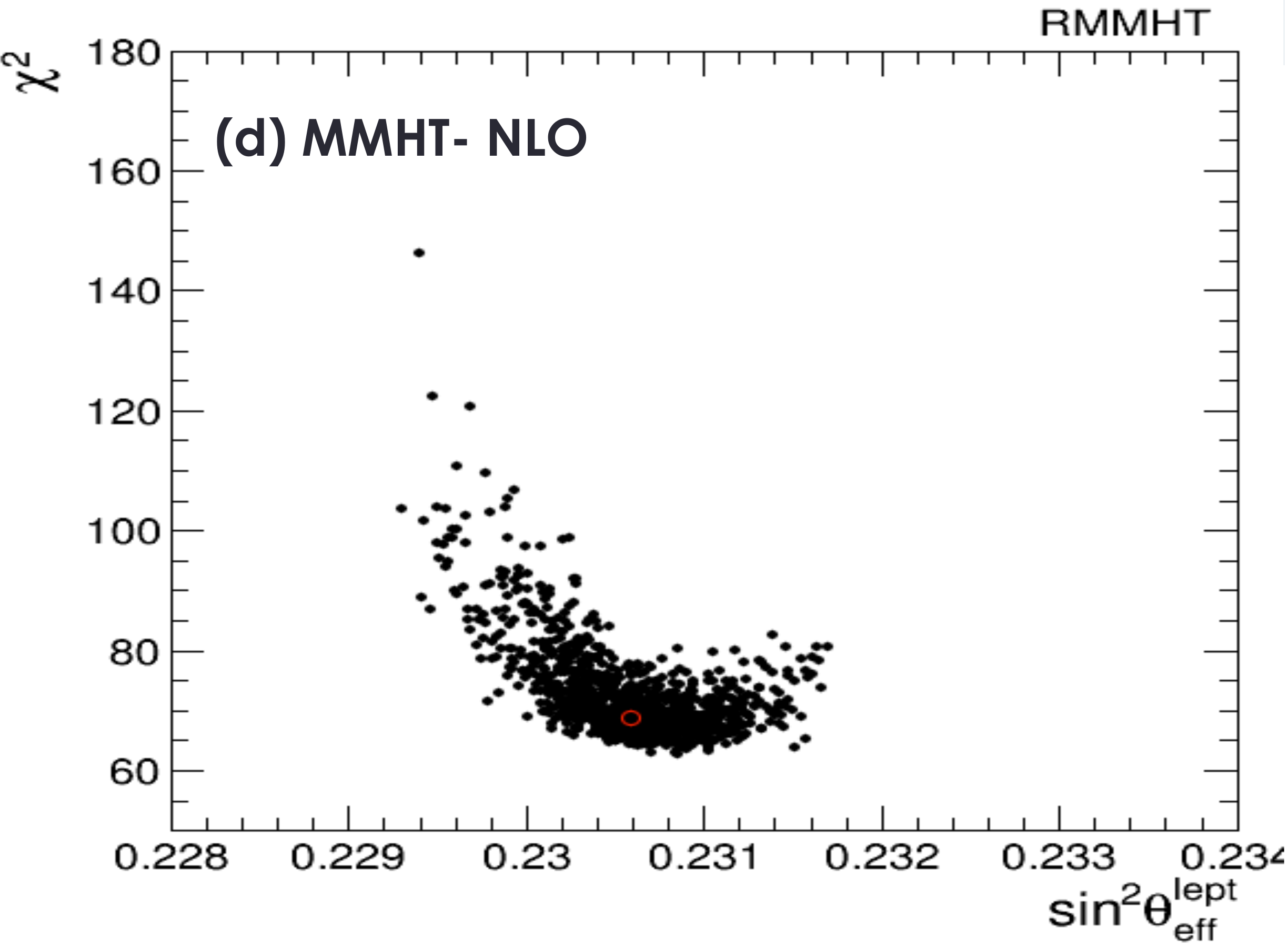}}
\caption{ Scatter plots of $\chi^{2}_{Afb}$ values versus $\sin^2\theta_{eff}$ 
for one of the 64  LHC pseudo experiments. 
  Here templates are  generated with 1000 replicas for   
(a)   \textsc{nnpdf} 3.0(\textsc{nlo}) (b) \textsc{CT10}(\textsc{nlo}), (c) \textsc{CT14}(\textsc{nlo}),  and
(d) \textsc{MMHT}(\textsc{nlo}).
 The number of degrees of freedom is 71 (=$6\times 12-1$).
The pseudo data are  generated 
with \textsc{powheg} with  the default  \textsc{nnpdf} 3.0 (\textsc{nlo}) PDF and $\sin^2\theta_{eff}$=0.23120.
 (6.7 M  dimuon events with  CMS-like detector acceptance cuts).
  } 
 \label{fig_9}
\end{figure}
\begin{figure}[ht]
\includegraphics[width=8.5 cm, height=8 cm]{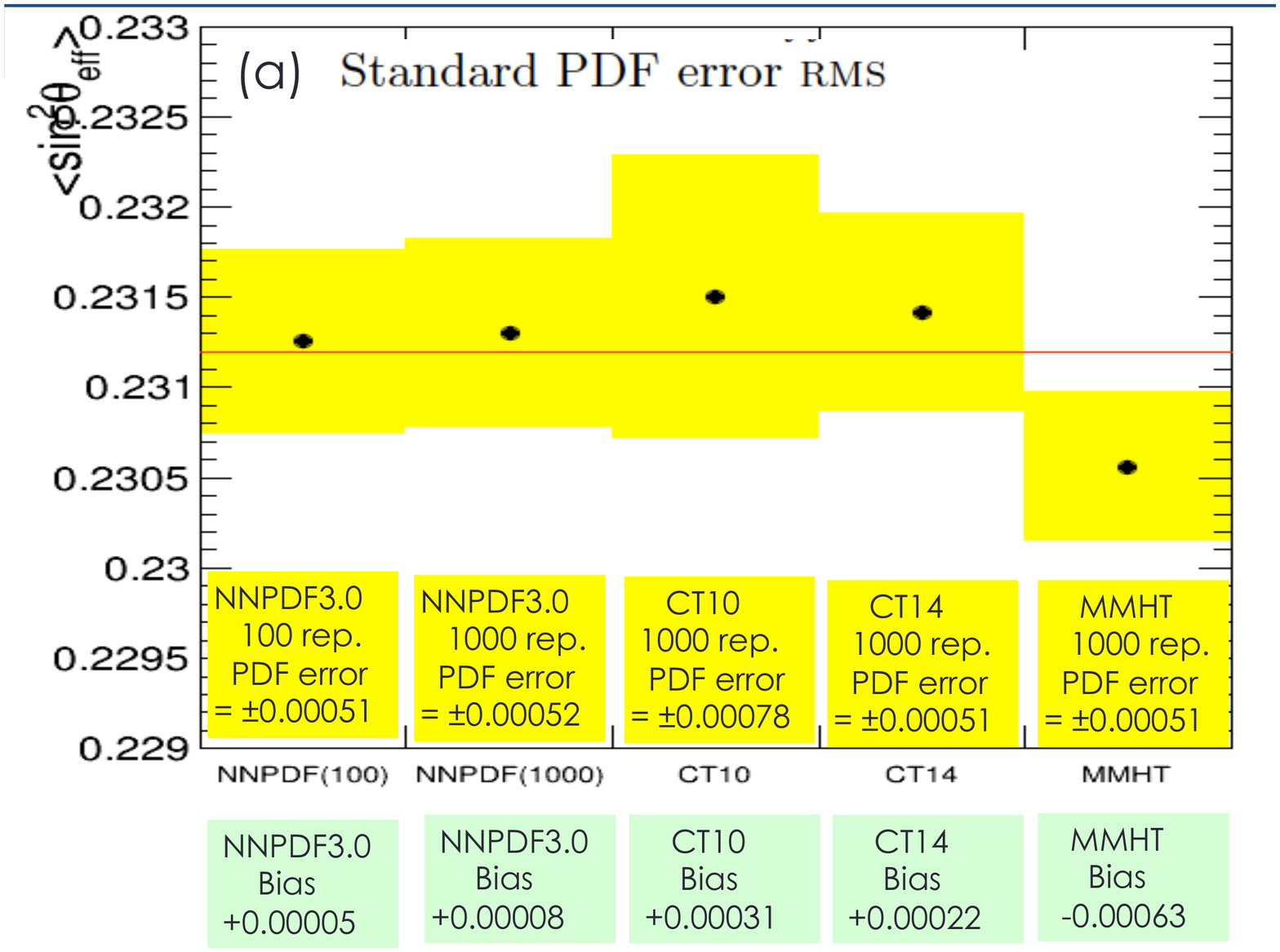}
\includegraphics[width=8.5 cm, height=8 cm]{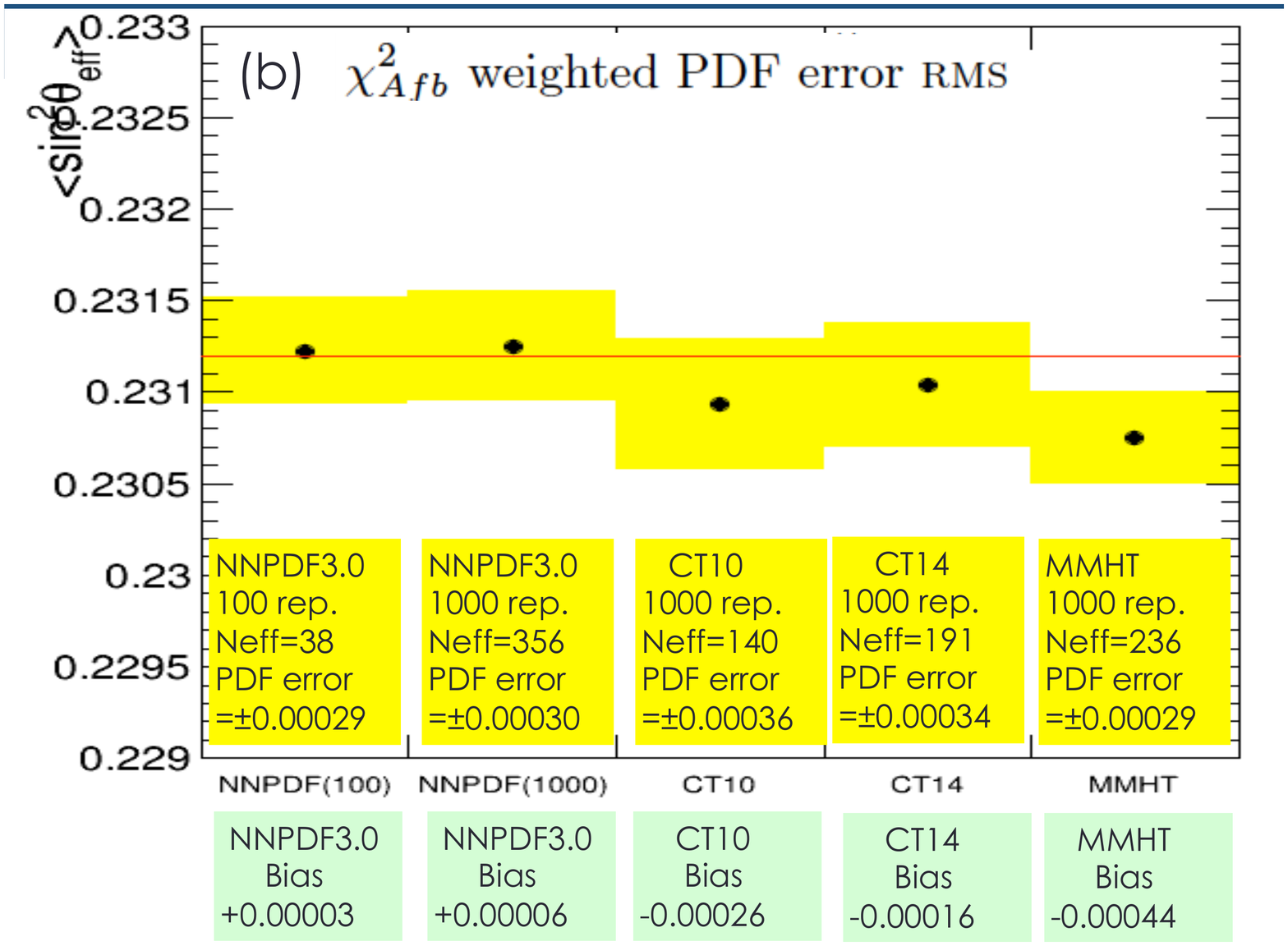}
\caption{  The average of the results from the analyses of
the  64  LHC pseudo experiments.
Each pseudo experiment is  analyzed with 100 NNPDF3.0 templates,  1000 NNPDF3.0 templates, 
1000 CT10 templates and 1000 MHHT templates.
The pseudo data for each experiment are   generated  by the \textsc{powheg}  MC with  the default  \textsc{nnpdf} 3.0 (\textsc{nlo}) PDF and $\sin^2\theta_{eff}$=0.23120. 
(a) Analysis using the standard mean and RMS of the  $\sin^2\theta_{eff}$  values extracted with  each PDF set.
(b) Analysis using the $\chi^{2}_{Afb}$ weighted  mean and RMS of  the  $\sin^2\theta_{eff}$  values extracted with  each PDF set.} 
 \label{fig_10}
\end{figure}
\begin{table}[htb]
\caption {Values of  $\sin^2\theta_W$ 
 with statistical errors and  PDF uncertainties  
 expected for a
 15 fb$^{-1}$ Drell-Yan dimuon sample at the LHC
 (at 8 TeV).
 The pseudo data are generated  by 
 the \textsc{powheg}  MC generator  with the  default  \textsc{nnpdf} 3.0 (\textsc{nlo}) PDF, and  
$\sin^2\theta_{eff}$=0.23120.
   The  PDF uncertainty for a standard  analysis is compared to the PDF
uncertainty for an analysis with both $\chi^{2}_{Afb}$ weighting
and $\chi^2_{Afb}+\chi^2_{Wasym}$ weighting.}
    \begin{center}
\begin{tabular}{|c||c|c||}
\hline
LHC~CMS~like	                  & input	\\
   Pseudo-Experiment&	\textsc{powheg} 	\\
{{LHC~15~fb$^{-1}$}} ~8~TeV  & Default    \\
$6.7M~\mu^+\mu^-$  &	{\textsc{nnpdf} 3.0 (\textsc{nlo})} 	\\
reconstructed~events &	(261000) 	\\
\hline
$\sin^2\theta_{eff}$ input &	0.23120 	\\
\hline\hline
statistical~error 	&	$\pm$0.00050		\\
$\Delta\sin^2\theta_{eff} $		&				\\
CT10~PDF~error & $\pm$ 0.00080\\
\hline \hline
Analysis replicas	&100		\\ 
\textsc{nnpdf}  set 	& \textsc{nnpdf} 3.0 (\textsc{nlo})				\\ 
Templates	     	&	\textsc{powheg} 	\\
		&				\\ 
		Average  method &	$N_{eff}=100$			\\
extracted ~$\sin^2\theta_{eff}$		&	0.23121	\\ 
{Standard PDF~error~\textsc{rms} }   	&		$\pm$0.00051	\\
(uncertainty in PDF error)&~(~0.00004) \\
		&		\\ 
$\chi^2_{Afb}$ weighting		&	$N_{eff}=37$				\\
extracted ~$\sin^2\theta_W$		&	0.23119			\\
{ $\chi^2_{Afb}$ weighted~PDF~error \textsc{rms} }		&		$\pm$0.00029		\\
(uncertainty in PDF error)&~(~0.00003) \\
&				\\ 
{$\chi^2_{Afb}$+$\chi^2_{Wsym}$~weighting }		&	$N_{eff}=15$				\\
extracted ~$\sin^2\theta_W$		&	0.23122			\\
{Weighted~PDF~error \textsc{rms} }		&		$\pm$0.00026		\\
(uncertainty in PDF error)&~(~0.00005) \\
%
\hline\hline
\end{tabular}
\label{table_3}
    \end{center}
\end{table}
\begin{table}[htb]
\caption {Expected statistical and weighted 
PDFs uncertainties  in  the measurements of  $ \sin^2\theta_W$ 
and $ M_W^{indirect}$  with a CMS like detector 
for two samples. 
(a) A total of 15M reconstructed  dilepton events 
(8.2 M $\mu^+\mu^-$ and 6.8M  $e^+e^-$) 
in a CMS like detector. This is
 is  similar to the  existing 19 fb$^{-1}$
CMS data sample  at 8 TeV.
(b) 120M reconstructed $\mu^+\mu^-$ events,
which the  sample expected  for  a CMS like detector with 
200 fb$^{-1}$  at 13-14 TeV}
    \begin{center}
\begin{tabular}{|c||c|c|c||}
\hline
CMS~like~detector&	 2016 & 2017-18\\
&	 sample  & sample\\
\hline
Energy	&	8~TeV 	&  13-14~TeV		\\
\hline
Number of  		& 8.2M $\mu^+\mu^-$	& 120M $\mu^+\mu^-$\\
  reconstructed events &	6.8M $e^+e^-$	&	-	\\
\hline
$\Delta \sin^2\theta_W$&  &\\
 Statistical~ error &	$\pm$ 0.00034& $\pm$ 0.00011 	\\
  Weighted~PDF~ error  &	$\pm$ 0.00022& $\pm$ 0.00014	\\
    (Stat+PDF)~error &	$\pm$ 0.00040&$\pm$ 0.00018 	\\
 \hline
 \hline
$\Delta M_W^{indirect}$& MeV & MeV \\
 Statistical~ error  &	$\pm$17&  $\pm$5 	\\
  weighted ~PDF~ error &	$\pm$11& $\pm$7	\\
    (Stat+PDF)~ error &$\pm$20  & $\pm$9	\\
\hline\hline
\end{tabular}
\label{table_4}
    \end{center}
\end{table}

  For each pseudo experiment we  find   the mean value and PDF uncertainty of  $\sin^2\theta_{eff}$ 
  from   the average and \textsc{rms}  of the $\sin^2\theta_{eff}$  for the  100 PDF replicas. 
   The average and \textsc{rms}  values are done in  three ways:
   \begin{enumerate}
\item       Using the standard  average and \textsc{rms}  of the  $\sin^2\theta_{eff}$ fit  values.
              This analysis results in a standard  PDF uncertainty of $\pm0.00051$  with 100 replicas.
 \item   Using  the  $\chi^{2}_{Afb}$ values  of the fits to $A_{FB} (M,y)$  to form  a weighted average and 
     weighted \textsc{rms}  of the  $\sin^2\theta_{eff}$  values.
        This analysis results in a  PDF uncertainty of $\pm0.00029$  with 37 effective  replicas.
  \item  Using  the combined  $\chi^{2}_{Afb}$+$\chi^2_{Wasym}$ 
        for the fits to Drell-Yan $A_{FB} (M,y)$ pseudo data  and the fits to the 
 W lepton decay asymmetry pseudo data   to form the weighted average and weighted \textsc{rms}  
  of the  $\sin^2\theta_{eff}$  values.
   This analysis results in a  PDF uncertainty of $\pm0.00026$  with 15 effective  replicas.
  \end{enumerate}
  \subsection{Studies with 1000 replicas}   
  As shown in Table~\ref{table_3},  the number of effective PDF replicas is reduced  to 15 when we
  apply constraints from both $\chi^{2}_{Afb}$ and   $\chi^2_{Wasym}$.  The PDF uncertainty
  is reduced to $\pm0.00026$.  The uncertainty in the estimate of the PDF uncertainty is $\pm$0.00005. 
  If we start with 1000 PDF replicas, the number of effective PDF replicas is $\approx$150,
  and the uncertainty in the estimate of the PDF uncertainty is reduced to  $\pm$0.00002.
  Therefore, the analysis is somewhat  more robust if we start with  1000 PDF replicas. 
 
  Fig.~\ref{fig_9} shows scatter plots of $\chi^{2}_{Afb}$ values versus $\sin^2\theta_{eff}$ 
for one of the 64  LHC pseudo experiments. 
  Here templates are  generated with 1000 replicas for   
(a)   \textsc{nnpdf} 3.0(\textsc{nlo}) PDF set (b) \textsc{CT10}(\textsc{nlo}) PDF set, (c) \textsc{CT14}(\textsc{nlo}) PDF set   and
(d) \textsc{MMHT}(\textsc{nlo}) PDF set. 
 The number of degrees of freedom is 71 (=$6\times 12-1$). The pseudo data are  generated 
with \textsc{powheg} with  the default  \textsc{nnpdf} 3.0 (\textsc{nlo}) PDF and $\sin^2\theta_{eff}$=0.23120.
 (6.7 M  dimuon events with  CMS-like detector acceptance cuts).

  In order to reduce the statistical error and investigate the PDF uncertainties, we take the
  average of 64 pseudo experiments.  The statistical error in the average
  of the  64  $\sin^2\theta_{eff}$  measurements is $\pm$0.00007 (=0.00052/8).
  Fig.~\ref{fig_10}  shows the average of the results from the  analyses of
all  64  LHC pseudo experiments with templates 
generated with 100 NNPDF3.0 replicas,  1000 NNPDF3.0 replicas, 
1000 CT10 replicas and 1000 MHHT replicas.  The standard mean and RMS (=PDF uncertainty)  of the  $\sin^2\theta_{eff}$  values extracted with  each PDF set are shown in  Fig.~\ref{fig_10}(a). The  $\chi^{2}_{Afb}$ weighted  mean and RMS(=PDF uncertainty)  of  the  $\sin^2\theta_{eff}$  values extracted with  each PDF set are shown in  Fig.~\ref{fig_10}(b).  

As expected, since the pseudo data are  generated 
with \textsc{powheg} with  the default  \textsc{nnpdf} 3.0 (\textsc{nlo}) PDF, the input
value of $\sin^2\theta_{eff}$=0.23120  is extracted with no bias when the pseudo data
are analyzed using  templates generated with either 100 or 1000 \textsc{nnpdf} 3.0 (\textsc{nlo}) replicas.
The PDF uncertainty is  reduced from $\pm$0.00052 to $\pm$0.00030 when  $\chi^{2}_{Afb}$ weighted  mean and RMS are used.

 The CT10 PDFs are less precise because they do not incorporate any LHC data. Consequently, 
 the uncertainties with CT10 PDFs are larger. The CT10  PDF  uncertainty is  reduced from $\pm$0.00078 to $\pm$0.00036
when  $\chi^{2}_{Afb}$ weighted  mean and RMS are used. Similarly, the  bias with CT10 is reduced from +0.00031 to -0.00026 which is within the  reduced  PDF uncertainty.
 The CT14 PDFs and MMHT PDFs incorporate LHC data in the fits. 
 The PDF uncertainties with CT14 are  reduced from $\pm$0.00051 to $\pm$0.00034 
when  $\chi^{2}_{Afb}$ weighted  mean and RMS are used. Similarly, the  bias with CT14  is reduced from +0.00022 to -0.00016, which is within the reduced PDF uncertainty.
  The PDF uncertainties with MMHT are  reduced from $\pm$0.00051 to $\pm$0.00029 with   $\chi^{2}_{Afb}$ weighted  mean and RMS. Here, the  bias with MMHT  is reduced from -0.00063 to -0.00044, but it is still larger than the PDF uncertainty. 
 
 As shown in   Fig.~\ref{fig_9}, the $A_{fb}$ analysis of the pseudo data illustrates that MMHT PDF set is not fully consistent with the  NNPDF or with  CT14 PDF set.   A similar study with actual  $A_{fb}$  data at 8 TeV would be a first step in the investigation of the origin of the differences between the various PDF sets.

%
 %
  \section{Conclusion}
 We show that measurements of the  Drell-Yan forward-backward charge asymmetry ($A_{FB}(M,y)$)  at hadron colliders 
 provide a new powerful tool to  reduce the  PDF uncertainties  in the measurement of electroweak
 parameters.

Table~\ref{table_4} summarizes the analysis for  two samples.
The first  (labeled 2016) is  a sample of 8.2M  $\mu^+\mu^-$  and 6.8M  $e^+e^-$  reconstructed events
     (with $M_{ll}>$~50 GeV) corresponding to an  integrated luminosity of 19 fb$^{-1}$  for a CMS like detector at 8 TeV.    This sample is  similar to the  existing 19 fb$^{-1}$
CMS data sample  at 8 TeV.
  The  statistical error in  the measurement  of $\sin^2\theta_{eff}$ for this sample
  is expected to be $\pm$ 0.00034, and the weighted PDF uncertainty is expected to be  $\pm$ 0.00022. These are  equivalent to a statistical  error of $\pm$17 MeV  and a weighted PDF uncertainty of $\pm$11 MeV  in the indirect measurement of $M_W$.    
 
With the larger number of $\mu^+\mu^-$ events expected to be collected at 13-14 TeV, both the statistical errors and the weighted  PDF uncertainties are  expected to be smaller.  About 120M reconstructed   $\mu^+\mu^-$   events  (with $M_{\mu\mu}>$~50 GeV) are expected  in a CMS like detector for an  integrated luminosity of 200 fb$^{-1}$  at 13-14 TeV.   For this sample (labeled 2017-18), as shown in the second column of Table~\ref{table_4},  the expected statistical error in the indirect measurement of $M_W$  is 5 MeV, and the weighted PDF uncertainty is $\pm$7 MeV.  These  expected errors  are smaller than the uncertainties in the most recent direct
 measurements of $M_W$.
 
 %
 
  \section{Acknowledgements}
  This work was support by the US Department of Energy, office of High Energy Physics under
  grant number DE-SC-0008475.

\end{document}